\documentclass[aps,prd,twocolumn,showpacs,superscriptaddress,groupedaddress,nofootinbib]{revtex4}
\usepackage{xcolor}
\usepackage{graphicx,color,bm}

\usepackage{amsmath,amssymb}
\usepackage{epstopdf}
\usepackage{ulem}
\usepackage{array}
\usepackage{verbatim}
\usepackage[utf8]{inputenc}
\usepackage{rotating}
\newcolumntype{K}[1]{>{\centering\arraybackslash}m{#1}}

\usepackage[colorlinks,linkcolor=blue,citecolor=blue]{hyperref}

\newcommand{\orcid}[1]{\href{https://orcid.org/#1}{\,\includegraphics[width=8px]{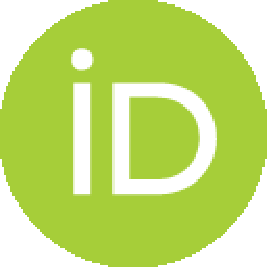}}}

\usepackage{mathtools}

\usepackage{tabularx}

\begin{document}

\title{Cosmic curvature on large-scale structures with homogeneous dark energy}

\author{Bikash R. Dinda \orcid{0000-0001-5432-667X}}
\email{bikashrdinda@gmail.com}
\affiliation{Department of Physical Sciences, Indian Institute of Science Education and Research Kolkata, Mohanpur, Nadia, West Bengal 741246, India.}
\affiliation{Department of Physics $\&$ Astronomy, University of the Western Cape, Cape Town, 7535, South Africa.}


\begin{abstract}
This study explores the impact of cosmic curvature on structure formation through general relativistic first-order perturbation theory. We analyze continuity and Euler equations, incorporating cosmic curvature into Einstein equations. Emphasizing late-time dynamics, we investigate matter density contrast evolution in the presence of cosmic curvature, with a specific focus on sub-hubble scales. Solving the evolution equation, we conduct data analysis using cosmic chronometers, baryon acoustic oscillations from Dark Energy Spectroscopic Instrument Data Release 2 (DESI DR2) (including $r_d$ constraint from CMB), type Ia supernova observations from Pantheon+ sample, logarithmic growth rate $f$ data and $f\sigma_8$ data. While constraints on some parameters remain consistent, inclusion of cosmic curvature losens constraints on $\Omega_{\rm m0}$ and $\sigma_{\rm 8,0}$ in $\Lambda$CDM and $w_0$CDM models. The non-phantom behavior of dark energy proves more favorable in $w_0$CDM model, whereas in CPL model, there is certain evidence for non-phantom behaviour at lower redshift and phantom behaviour at higher redshift. Interestingly, we find certain evidence for non-zero cosmic curvature in $\Lambda$CDM model, but no evidence for it in both $w_0$CDM and CPL models, which indirectly proves the non-degeneracy between dynamical dark energy and cosmic curvature.
\end{abstract}


\maketitle
\date{\today}

\section{Introduction}

The investigation into the late-time cosmic dynamics by observations like type Ia supernovae \cite{SupernovaCosmologyProject:1997zqe,SupernovaSearchTeam:1998fmf,SupernovaCosmologyProject:1998vns,2011NatPh...7Q.833W,Linden2009CosmologicalPE,Camarena:2019rmj,Pan-STARRS1:2017jku,Camlibel:2020xbn}, cosmic microwave background \cite{Planck:2013pxb,Planck:2015fie,Planck:2018vyg}, baryon acoustic oscillations \cite{BOSS:2016wmc,eBOSS:2020yzd,Hou:2020rse}, and cosmic chronometers \cite{Jimenez:2001gg,Pinho:2018unz,Cao:2023eja}, has strongly established the accelerating expansion of our Universe. This cosmic acceleration, however, can not be explained within the standard matter or dark matter paradigms governed by general relativity. Consequently, two popular theoretical aspects have emerged to explain this late-time cosmic acceleration: the existence of dark energy \cite{Peebles:2002gy,Copeland:2006wr,Yoo:2012ug,Lonappan:2017lzt,Dinda:2017swh,Dinda:2018uwm} or the modifications to the general theory of relativity on cosmic scales \cite{Clifton:2011jh,Koyama:2015vza,Tsujikawa:2010zza,Joyce:2016vqv,Dinda:2017lpz,Dinda:2018eyt,Zhang:2020qkd,Dinda:2022ixi,Bassi:2023vaq,Nojiri:2010wj,Nojiri:2017ncd,Bamba:2012cp,Lee:2022cyh}. While dark energy being a constituent of the Universe with large negative pressure, capable of driving the cosmic acceleration, alternative theories of modified gravity account for the observed cosmic acceleration without introducing any dark energy component.

Among different dark energy models, arguably the $\Lambda$CDM model stands out as the most successful \cite{Carroll:2000fy}. In this model, a cosmological constant ($\Lambda$), with a constant equation of state, with value $-1$ acts as a dark energy component. Despite its success, the $\Lambda$CDM model has some shortcomings. For example, from the theoretical perspective, issues like cosmic coincidence and fine-tuning problems \cite{Zlatev:1998tr,Sahni:1999gb,Velten:2014nra,Malquarti:2003hn} are present. Moreover, observational tensions, including the Hubble tension \cite{DiValentino:2021izs,Krishnan:2021dyb,Vagnozzi:2019ezj,Dinda:2021ffa} and $\sigma_8$ (a quantity related to matter growth) tension \cite{DiValentino:2020vvd,Abdalla:2022yfr,Douspis:2018xlj,Bhattacharyya:2018fwb} corresponding to the discrepencies in measured values of $H_0$ (present value of the Hubble parameter) and $\sigma_{\rm 8,0}$ (present value of $\sigma_8$) between early and late time observations. This necessitates studies of models beyond $\Lambda$CDM.

The standard $\Lambda$CDM model assumes that the 3-space of the four-dimensional space-time is flat \cite{Planck:2013pxb,Planck:2015fie,Planck:2018vyg}. However, it is important to incorporate the influence of cosmic curvature into our analysis to study its impact. While literature acknowledges prior investigations exploring the prospect of non-zero cosmic curvature in the Universe \cite{Handley:2019tkm,Desgrange:2019npu,Coley:2019yov,DiValentino:2020srs,Moresco:2016nqq,Wei:2016xti,Wang:2017lri,Cheng-Zong:2019iau,Yang:2020bpv,Li_2019,Wang:2019yob,Dinda:2023svr,Yu:2016gmd,Wei_2020,Liu:2020pfa,Mukherjee:2022ujw,Shi:2023kvu,Vagnozzi:2020dfn,Dhawan:2021mel,Yang:2022kho,Jesus:2019jvk}, these studies predominantly focus on the background dynamics of cosmic expansion, there are very few studies to examine the impact of cosmic curvature on the evolution of perturbations \cite{Bel:2022iuf,Eingorn:2019spk,Abbassi:2020drc}. The inclusion of cosmic curvature affects the measurements of key parameters like $\Omega_{\rm m0}$ (present value of matter energy density parameter) and $\sigma_{\rm 8,0}$.

Regarding the dynamical dark energy models beyond the $\Lambda$CDM, there are potential degeneracies between the equation of state for dark energy and cosmic curvature. This fact necessitates a comprehensive exploration of alternative dark energy models \cite{Clarkson:2007bc,Gao_2020,Wang:2007mza,Gong:2007wx,Alonso:2023oro} with the presence of cosmic curvature. Specifically, we consider the $w_0$CDM and the Chevallier-Polarski-Linder (CPL) \cite{Chevallier:2000qy,Linder:2002et} model. The $w_0$CDM model features a dark energy equation of state that remains static over time but can take various values, including the cosmological constant's value of $-1$. On the other hand, the CPL model introduces dynamical variations in the equation of state of dark energy. By adopting these models, we aim to scrutinize the potential degeneracies between the equation of state for dark energy and cosmic curvature and the implications of these in the cosmological dynamics.

This paper is structured as follows: In Sec.~\ref{sec-DE_models}, we briefly discuss three popular dark energy parametrizations; In Sec.~\ref{sec-late_time_bkg}, we mention a brief overview of the background expansion dynamics; In Sec.~\ref{sec-late_subHub_nodepert}, we discuss the evolution of the matter inhomogeneity on sub-Hubble scales with the assumption that the considered dark energy models do not cluster on sub-Hubble scales; the detailed derivations are mentioned in Appendices~\ref{sec-metric} to~\ref{sec-late_time_sub-Hubble}; The four observational datasets considered in this analysis are briefly outlined in Sec.~\ref{sec-data}; In Sec.~\ref{sec-result} we present our results; Finally, in Sec.~\ref{sec-conclusion} we make a conclusion.

\section{Dark energy parametrizations}
\label{sec-DE_models}

In this study, we focus on three dark energy parametrizations for the equation of state of dark energy $w_Q$: the $\Lambda$CDM, the $w_0$CDM, and the Chevallier-Polarski-Linder (CPL; also called as $w_0w_a$CDM) \cite{Chevallier:2000qy,Linder:2002et}. In these parametrizations, the equation of state for dark energy is expressed as

\begin{eqnarray}
w_Q \hspace{0.1 cm} (\text{$\Lambda$CDM}) &=& -1,
\label{eq:w_DE_LCDM} \\
w_Q \hspace{0.1 cm} (w_0\text{CDM}) &=& w_0,
\label{eq:w_DE_wCDM} \\
w_Q \hspace{0.1 cm} (\text{CPL}) &=& w_0+w_a \frac{z}{1+z},
\label{eq:w_DE_CPL}
\end{eqnarray}

\noindent
respectively, where $z$ is the redshift, $w_0$ (both for $w_0$CDM and CPL) and $w_a$ (only for CPL) are the two model parameters. Therefore, the $w_0$CDM model is a subset of the CPL model where $w_a=0$. The $\Lambda$CDM model is a subset of the CPL model where $w_0=-1$ and $w_a=0$, and it is also a subset of the $w_0$CDM model where $w_0=-1$.

The normalized dark energy density is defined as $f_Q=\frac{\bar{\rho}_Q(z)}{\bar{\rho}_Q(z=0)}$, where $\bar{\rho}_Q$ is the background energy density of the dark energy. This can be computed from $w_Q$ as

\begin{equation}
f_Q = \exp \left[ 3 \int_0^z \frac{1+w_Q(\tilde{z})}{1+\tilde{z}} d\tilde{z} \right] ,
\label{eq:f_DE}
\end{equation}

\noindent
where $\tilde{z}$ is the dummy variable for the redshift $z$. By substituting Eqs.\eqref{eq:w_DE_LCDM}, \eqref{eq:w_DE_wCDM}, and \eqref{eq:w_DE_CPL} into Eq.\eqref{eq:f_DE}, we obtain the expressions of $f_Q$ for the $\Lambda$CDM, the $w_0$CDM, and the CPL models

\begin{eqnarray}
f_Q \hspace{0.1 cm} (\text{$\Lambda$CDM}) &=& 1,
\label{eq:f_DE_LCDM} \\
f_Q \hspace{0.1 cm} (w_0\text{CDM}) &=& (1+z)^{3(1+w_0)},
\label{eq:f_DE_wCDM} \\
f_Q \hspace{0.1 cm} (\text{CPL}) &=& (1+z)^{3(1+w_0+w_a)} e^{ -3 w_a \frac{z}{1+z} },
\label{eq:f_DE_CPL}
\end{eqnarray}

\noindent
respectively.

\begin{figure*}
\centering
\includegraphics[width=0.49\textwidth]{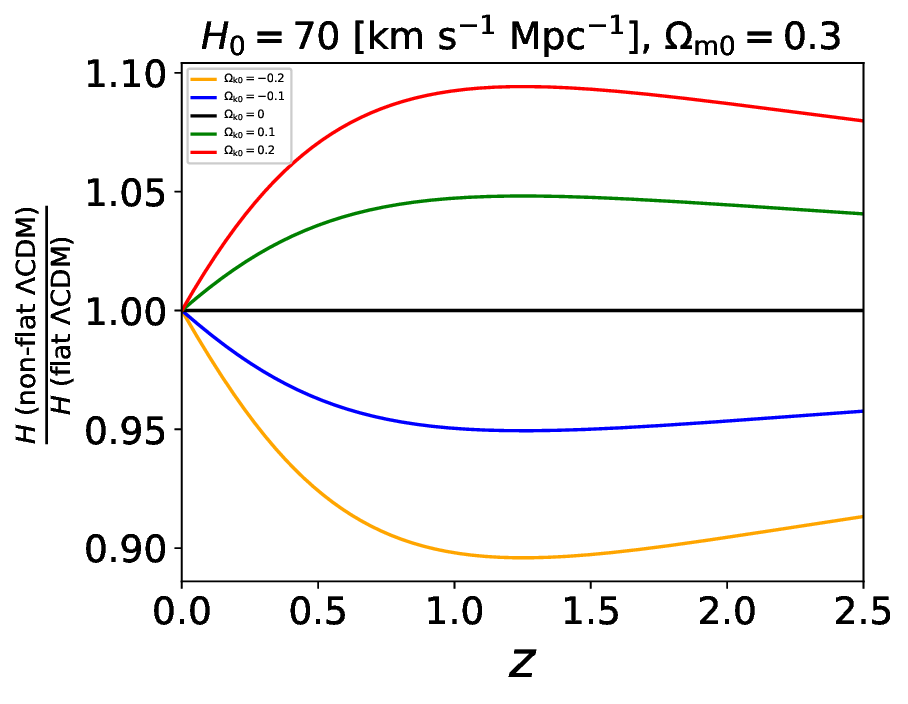}
\includegraphics[width=0.49\textwidth]{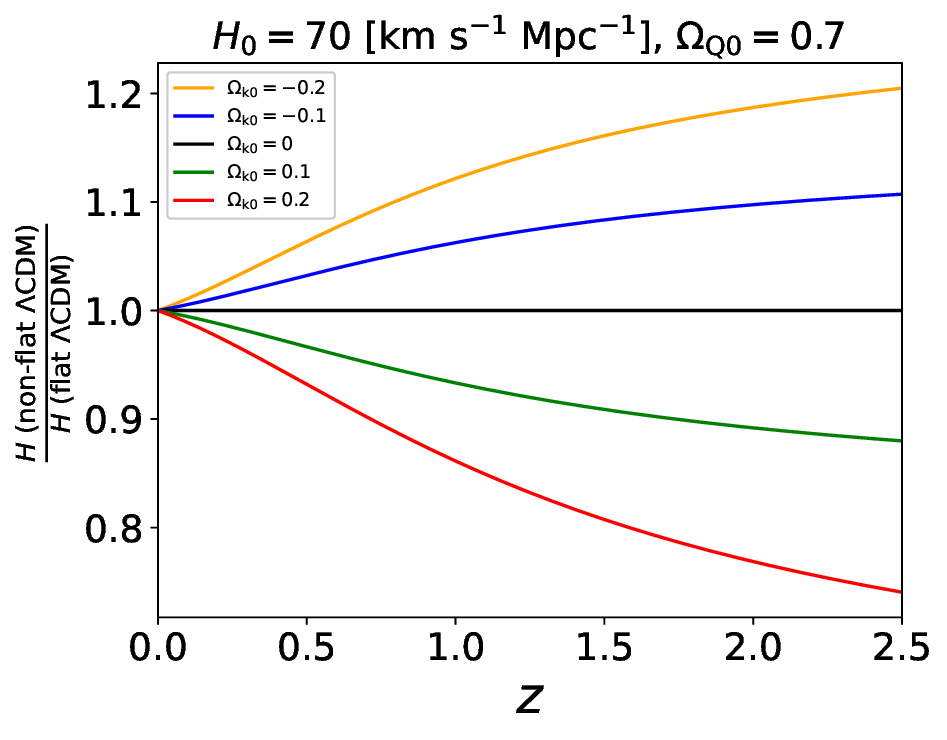}
\caption{
\label{fig:H_cmp_Ok0}
Comparison in $H$ between flat and non-flat $\Lambda$CDM model. The orange, blue, black, green, and red lines are for $\Omega_{\rm k0}$ values -0.2, -0.1, 0, 0.1, and 0.2 respectively. The left and right panels correspond to fixed $\Omega_{\rm m0}=0.3$ and fixed $\Omega_{\rm Q0}=0.7$ respectively.
}
\end{figure*}

\section{The late-time background}
\label{sec-late_time_bkg}

In the Friedmann-Lema\^itre-Robertson-Walker (FLRW) background, the normalized Hubble parameter $E$ (see the background Einstein equation \eqref{eq:Friedmann_1} in Appendix~\ref{sec-late_time_Einstein_equations} for more details) at late-time is given as

\begin{equation}
E^2 = \frac{H^2}{H_0^2} = \Omega_{\rm m0}(1+z)^{3}+\Omega_{\rm k0}(1+z)^{2} + \Omega_{\rm Q0} f_Q,
\label{eq:Esqr}
\end{equation}

\noindent
where $H$ is the Hubble parameter and $\Omega_{\rm Q0}$ is the present value of the dark energy density parameter $\Omega_Q$; $\Omega_{\rm k0}$ is given as $\Omega_{\rm k0}=-\frac{c^2\kappa}{H_0^2}$, where $\kappa$ is the spatial curvature of the space-time and $c$ is the speed of light in vacuum; $\kappa<0$, $\kappa=0$, and $\kappa>0$ correspond to the open, flat, and closed Universe respectively.

Note that, in Eq.\eqref{eq:Esqr}, we have neglected the contribution from radiation because radiation hardly has any contribution at late times.

We have an energy budget restriction condition at present given as

\begin{equation}
\Omega_{\rm m0}+\Omega_{\rm Q0}+\Omega_{\rm k0} = 1 .
\label{eq:budget_constraint}
\end{equation}

We can also compute other relevant background quantities such as $\Omega_m$ (matter energy density parameter), $\Omega_Q$, and $\Omega_k$ as

\begin{align}
\label{eq:Om_wrt_z_a}
\Omega_m &= \frac{8\pi G\bar{\rho}_m}{3H^2} = \frac{\Omega_{\rm m0} (1+z)^3}{E^2} = \frac{\Omega_{\rm m0} a^{-3}}{E^2} , \\
\label{eq:OQ_wrt_z_a}
\Omega_Q &= \frac{8\pi G\bar{\rho}_Q}{3H^2} = \frac{\Omega_{\rm Q0} f_Q}{E^2} , \\
\label{eq:Ok_wrt_z_a}
\Omega_k &= - \frac{c^2 \kappa}{a^2H^2} = \frac{\Omega_{\rm k0} (1+z)^2}{E^2}  = \frac{\Omega_{\rm k0} a^{-2}}{E^2},
\end{align}

\noindent
respectively, where $a$ is the scale factor and it is related to redshift as $a=\frac{1}{1+z}$; $G$ is the Newtonian gravitational constant; $\bar{\rho}_m$ is the energy density of matter. A restriction equation similar to Eq.\eqref{eq:budget_constraint} is also applicable at any given redshift given as $\Omega_m+\Omega_Q+\Omega_k=1$.

The cosmic curvature affects the total energy budgets of the Universe and hence the evolutionary background expansion. This can be seen through the Hubble parameter. To see this, we check how the Hubble parameter changes if we include cosmic curvature in the standard $\Lambda$CDM model. This has been plotted in Fig.~\ref{fig:H_cmp_Ok0}. For this plot, we have fixed $H_0=70$ km$~$s$^{-1}~$Mpc$^{-1}$ and and we vary $\Omega_{\rm k0}$. For left panel, we fix $\Omega_{\rm m0}=0.3$ and for right panel we fix $\Omega_{\rm Q0}=0.7$. The orange, blue, black, green, and red lines are for $\Omega_{\rm k0}$ values -0.2, -0.1, 0, 0.1, and 0.2 respectively. We can see that cosmic curvature has a significant effect on the background expansion: around 4 to 10 percent for $|\Omega_{\rm k0}|=0.1$ and $|\Omega_{\rm k0}|=0.2$ respectively, when $\Omega_{\rm m0}$ is fixed and up to 10 to 20 percent for $|\Omega_{\rm k0}|=0.1$ and $|\Omega_{\rm k0}|=0.2$ respectively, when $\Omega_{\rm Q0}$ is fixed. Also, the positive and negative curvatures have the opposite effects: (i) when $\Omega_{\rm m0}$ is fixed, positive $\Omega_{\rm k0}$ has a faster expansion rate compared to the flat counterpart whereas negative $\Omega_{\rm k0}$ has a slower expansion rate, and (ii) when $\Omega_{\rm Q0}$ is fixed, it's exactly the opposite.

\begin{figure*}
\centering
\includegraphics[width=0.49\textwidth]{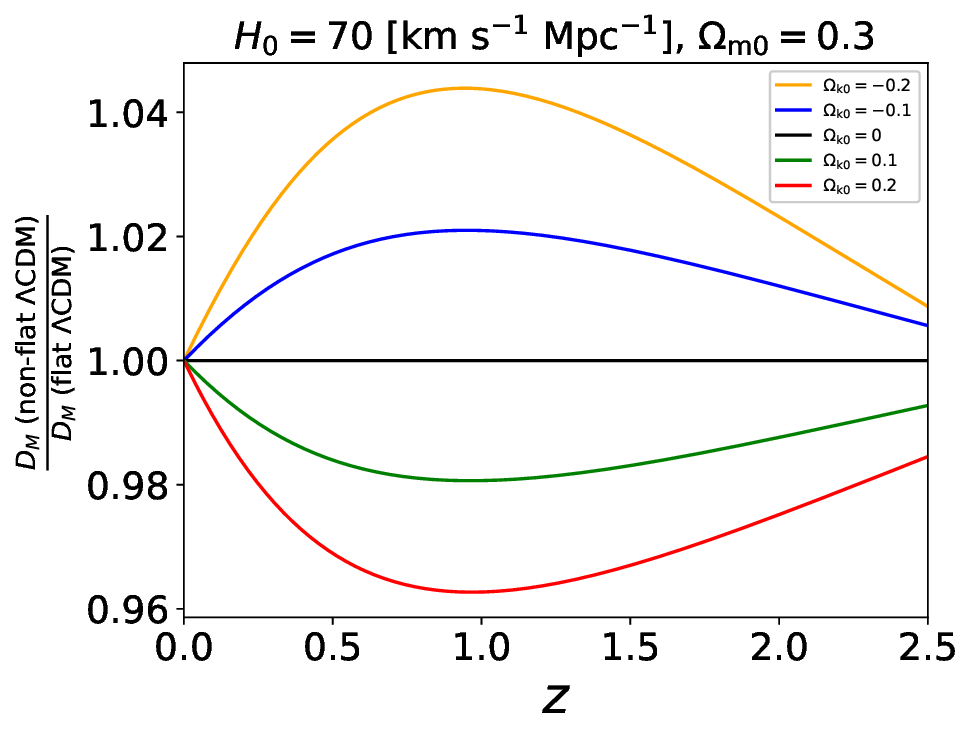}
\includegraphics[width=0.49\textwidth]{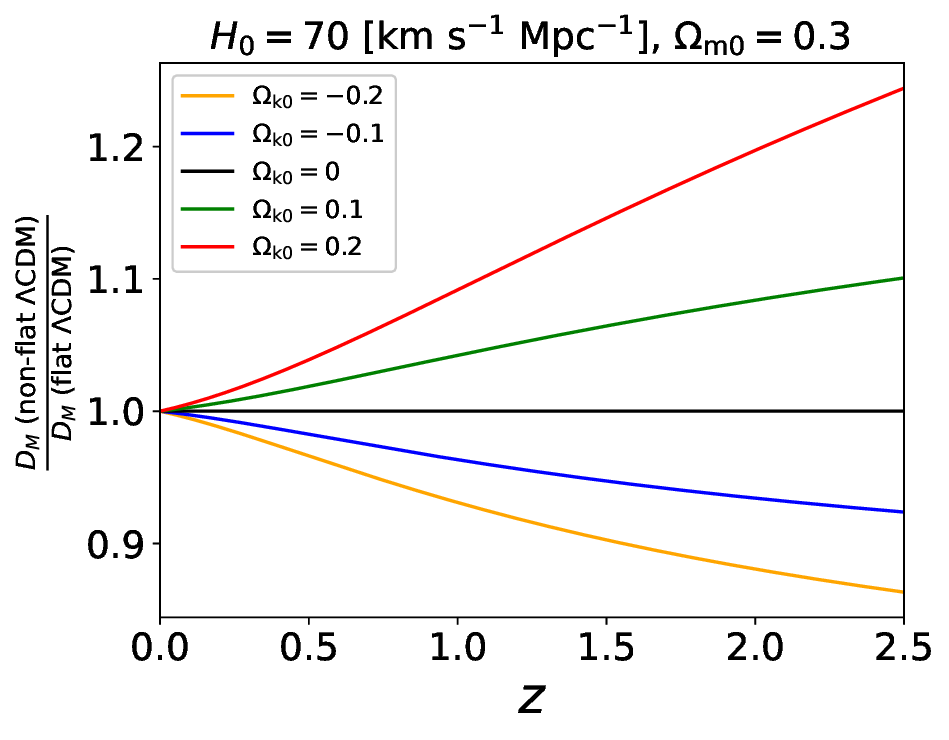}
\caption{
\label{fig:DM_Ok0}
Comparison in $D_M$ between flat and non-flat $\Lambda$CDM model. The orange, blue, black, green, and red lines are for $\Omega_{\rm k0}$ values -0.2, -0.1, 0, 0.1, and 0.2 respectively. The left and right panels correspond to fixed $\Omega_{\rm m0}=0.3$ and fixed $\Omega_{\rm Q0}=0.7$ respectively.
}
\end{figure*}

The transverse comoving distance $D_M$ in a curved background is given as
\begin{equation}
    D_M = \begin{cases}
    \frac{c}{H_0\sqrt{\Omega_{\rm k0}}} \sinh \left( \sqrt{\Omega_{\rm k0}} d_m \right), & \mbox{for } \Omega_{\rm k0}>0, \\
    \frac{c}{H_0} d_m, & \mbox{for } \Omega_{k0} = 0, \\
    \frac{c}{H_0\sqrt{|\Omega_{\rm k0}|}} \sin \left( \sqrt{|\Omega_{\rm k0}|} d_m \right), & \mbox{for } \Omega_{\rm k0}<0, \\
    \end{cases}
    \label{eq:DM}
\end{equation}
where $d_m$ is given as
\begin{equation}
d_m = \int_0^z \frac{d\tilde{z}}{ E(\tilde{z}) } .
\label{eq:H_to_dP}
\end{equation}

In Fig.~\ref{fig:DM_Ok0}, we plot $D_M$ similar to Fig.~\ref{fig:H_cmp_Ok0}. The left panel is for fixed $\Omega_{\rm m0}=0.3$ and the right panel is for fixed $\Omega_{\rm Q0}=0.7$. We can see the effect of cosmic curvature is around 4 to 10 percent for $|\Omega_{\rm k0}|=0.1$ and $|\Omega_{\rm k0}|=0.2$ respectively, when $\Omega_{\rm m0}$ is fixed. The deviations are up to 5, 10, 15 and 25 percent for $\Omega_{\rm k0}=-0.1$, $\Omega_{\rm k0}=-0.2$, $\Omega_{\rm k0}=0.1$ and $\Omega_{\rm k0}=0.2$ respectively, when $\Omega_{\rm Q0}$ is fixed. The signatures of the deviations are opposite between fixed $\Omega_{\rm m0}$ and fixed $\Omega_{\rm Q0}$. One more thing to notice here is that positive $\Omega_{\rm k0}$ has larger deviations (with same absolute value) compared to the negative $\Omega_{\rm k0}$, when $\Omega_{\rm Q0}$ is fixed.

\section{Linear matter perturbation at late-time: sub-Hubble approximation and homogeneous dark energy}
\label{sec-late_subHub_nodepert}

\subsection{Evolution of matter inhomogeneity}

In this study, we focus on the sub-Hubble approximation and homogeneous dark energy models, where dark energy has no fluctuations. In this case, the evolution equation for the matter density contrast\footnote{The matter density contrast $\delta_m$ is defined as $\rho_m=\bar{\rho}_m(1+\delta_m)$, where $\rho_m$ is the total (background+perturbed) energy density of matter.} becomes (see Eq.\eqref{eq:dm_double_dot_sub_Hubble_no_de_pert} in Appendix~\ref{sec-late_time_sub-Hubble})

\begin{equation}
\ddot{\delta}_m + 2 H \dot{\delta}_m - 4\pi G \bar{\rho}_m \delta_m = 0 ,
\label{eq:dm_eqn_sub_Hubble_no_de_pert}
\end{equation}

\noindent
where over-dot and double-dot represent the first and second derivatives w.r.t cosmic time. For a detailed derivation of the above differential equation see Appendices~\ref{sec-metric} to~\ref{sec-late_time_sub-Hubble}.

Replacing the cosmic time derivatives w.r.t the scale factor $a$ and using Eq.\eqref{eq:Om_wrt_z_a} for the expression of $\bar{\rho}_m$ w.r.t $\Omega_m$, Eq.\eqref{eq:dm_eqn_sub_Hubble_no_de_pert} can be rewritten as

\begin{equation}
\frac{d^2\delta_m}{da^2} + \frac{1}{a} \left(3+\frac{\dot{H}}{H^2}\right) \frac{d\delta_m}{da} - \frac{3}{2} \frac{\Omega_m}{a^2} \delta_m = 0 ,
\label{eq:dm_evolution_dimensionless}
\end{equation}

\noindent
where we have\footnote{This equation is derived from the combination of Friedmann's first and second equations.}

\begin{equation}
\frac{\dot{H}}{H^2} = \frac{\Omega_k}{2}-\frac{3}{2} \left(1+w_Q\Omega_Q\right) .
\label{eq:HdotbyHsqr}
\end{equation}

\subsection{Numerical computation of growth function}

The growth function $D_+$ is defined as

\begin{equation}
\delta_m = D_+ \delta_m^i ,
\label{eq:growth_function}
\end{equation}

\noindent
where $\delta_m^i$ is the initial value of $\delta_m$ at a chosen initial epoch. Since $D_+$ is proportional to $\delta_m$, it follows the same differential equation as in Eq.\eqref{eq:dm_evolution_dimensionless}. This differential equation is second-order and can be divided into two first-order differential equations to make an autonomous system of differential equations given as

\begin{eqnarray}
\dfrac{dX}{da} &=& Y , \nonumber\\
\dfrac{dY}{da} &=& AX+BY ,
\label{eq:autonomous_diff_eqns}
\end{eqnarray}

\noindent
where we have

\begin{eqnarray}
X &=& D_+ , \nonumber\\
Y &=& \dfrac{dD_+}{da} , \nonumber\\
A &=& \frac{3}{2} \frac{\Omega_m}{a^2} , \nonumber\\
B &=& - \frac{1}{a} \left(3+\frac{\dot{H}}{H^2}\right) .
\end{eqnarray}

\subsection{Initial conditions at matter-dominated era}

We consider initial conditions in the early matter-dominated era to solve the differential equation in Eq.\eqref{eq:dm_evolution_dimensionless} at an initial scale factor $a_i$. From now onwards, we denote the initial value of any quantity by the subscript "i". In the matter-dominated era, we approximately have $\Omega_m \simeq 1$, $\Omega_k \simeq 0$, and $\Omega_Q \simeq 0$. With these approximations, the evolution equation for $\delta_m$ during the matter-dominated era becomes

\begin{equation}
\frac{d^2\delta_m}{da^2} + \frac{3}{2a} \frac{d\delta_m}{da} - \frac{3}{2a^2} \delta_m \simeq 0 . ~  \text{(matter-dominated)}
\label{eq:dm_homo_MD}
\end{equation}

\noindent
There are two solutions to the above differential equation. One is proportional to $a^{-3/2}$, representing the decaying solution. The other is the growing solution, corresponding to $\delta_m \propto a$. And thus the first derivative i.e. $\dfrac{d\delta_m}{da}$ is approximately a constant at matter-dominated era. Using these two conditions for the growing mode solutions and using Eq.\eqref{eq:growth_function}, we set the initial conditions accordingly as

\begin{eqnarray}
X_i &=& 1, \nonumber\\
Y_i &=& \frac{1}{a_i} .
\label{eq:initial_conditions}
\end{eqnarray}

We do not consider arbitrary values of $a_i$ in the above equations, but we compute these values for which the matter energy density parameter $\Omega_m$ is the closest to $1$. For details, see \autoref{sec-ai}.

With the initial conditions, mentioned in Eq.\eqref{eq:initial_conditions}, we solve the system of autonomous differential equations in Eq.\eqref{eq:autonomous_diff_eqns} and find numerical solutions for $X$ and $Y$ as functions of $a$ and consequently as functions of $z$. Then we can find any other first-order perturbed quantities from these solutions. Note that, in Eq.\eqref{eq:autonomous_diff_eqns}, $A$ and $B$ are background quantities i.e. these are functions of $a$.

\begin{figure*}
\centering
\includegraphics[width=0.49\textwidth]{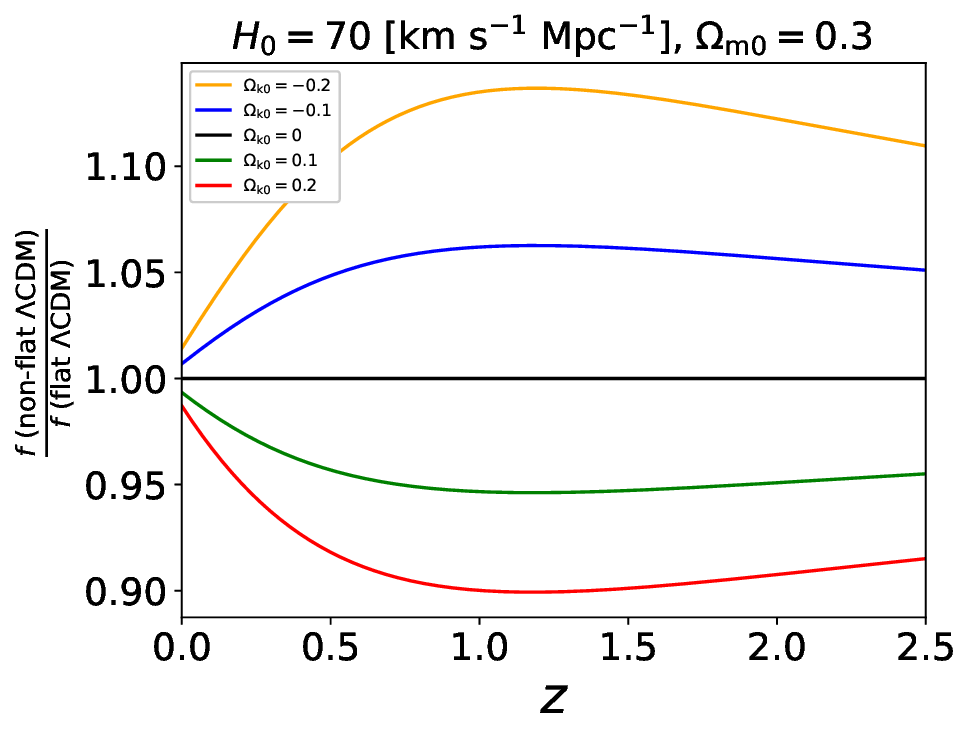}
\includegraphics[width=0.49\textwidth]{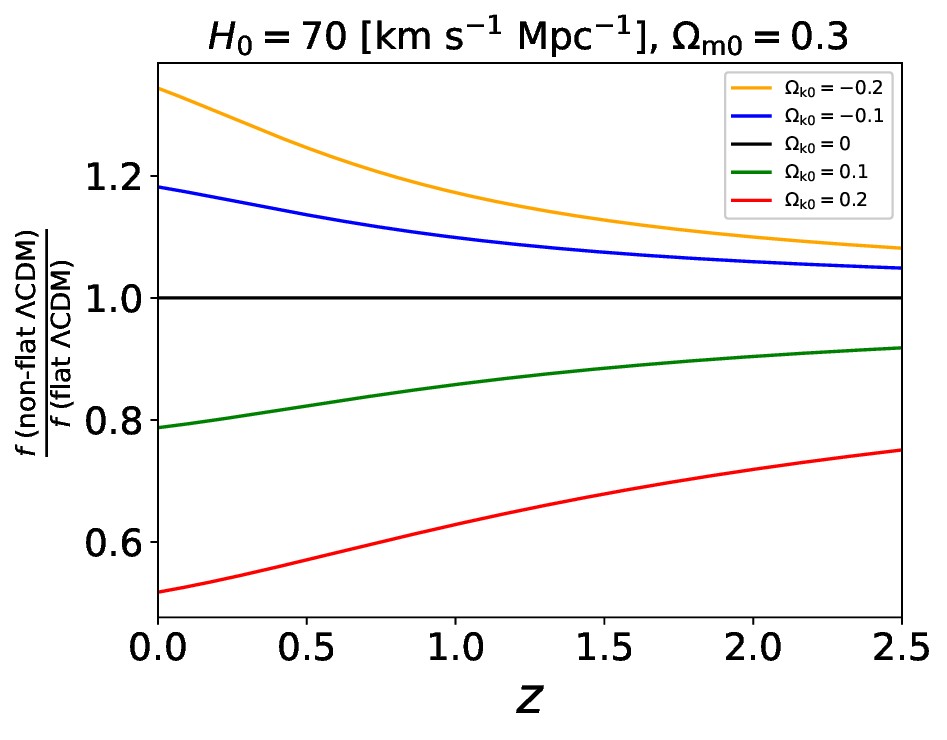}
\caption{
\label{fig:f_cmp_Ok0}
Comparison in $f$ between flat and non-flat $\Lambda$CDM model. The left and right panels correspond to fixed $\Omega_{\rm m0}=0.3$ and fixed $\Omega_{\rm Q0}=0.7$ respectively.
}
\end{figure*}

\begin{figure*}
\centering
\includegraphics[width=0.49\textwidth]{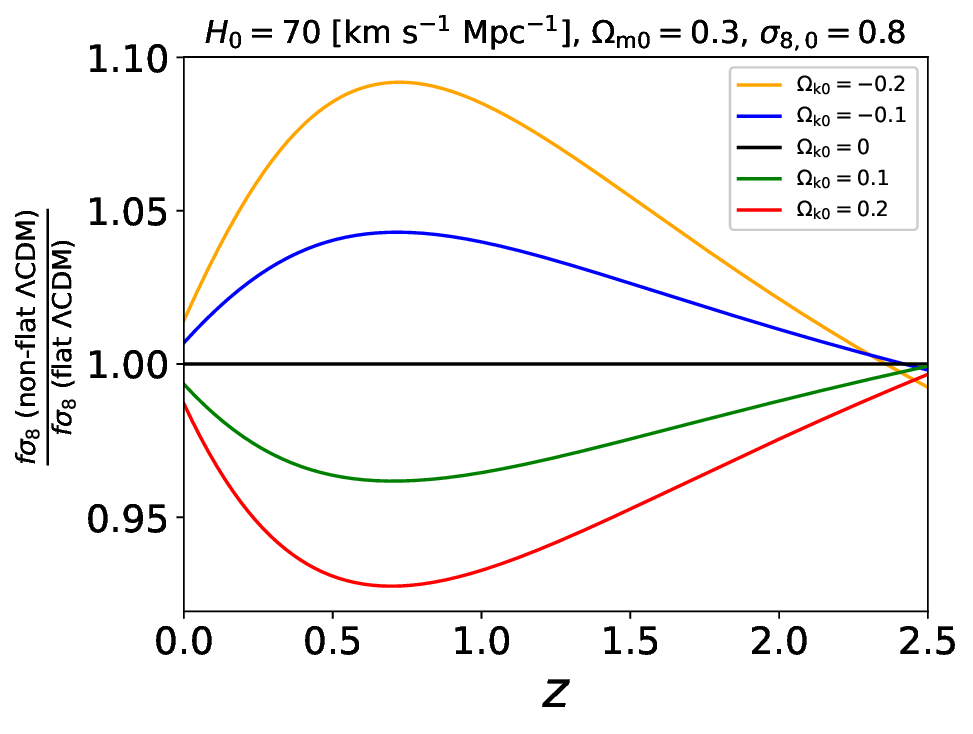}
\includegraphics[width=0.49\textwidth]{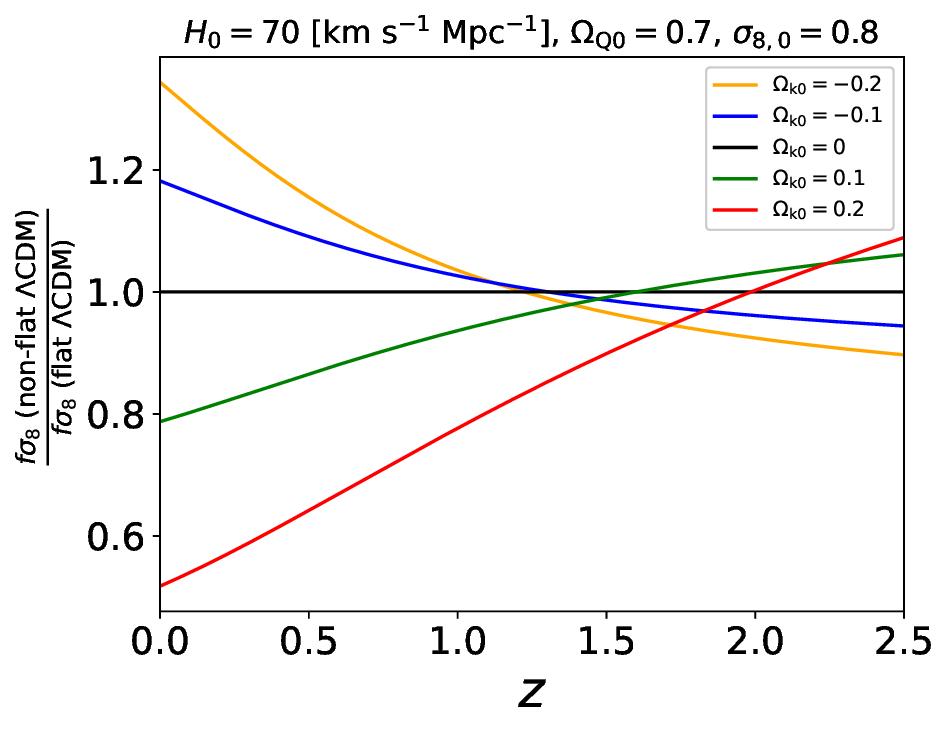}
\caption{
\label{fig:fsigma8_cmp_Ok0}
Comparison in $f\sigma_8$ between flat and non-flat $\Lambda$CDM model. The left and right panels correspond to fixed $\Omega_{\rm m0}=0.3$ and fixed $\Omega_{\rm Q0}=0.7$ respectively.
}
\end{figure*}

\subsection{$f$ and $\sigma_8$}

We are now turning our attention to two additional quantities. The first one is the logarithmic growth factor, defined as \cite{Huterer:2013xky}

\begin{equation}
f = \frac{d \ln D_{+} }{d \ln a} = \frac{aY}{X} .
\label{eq:growth_f}
\end{equation}

\noindent
In the sub-Hubble limit, within this framework, the normalization factor of the matter power spectrum, denoted as $\sigma_8$, retains its scale independence and can be represented as \cite{Pierpaoli:2000ip}

\begin{equation}
\sigma_8(z) = \sigma_{\rm 8,0} \frac{D_{+}(z)}{D_{+}(z=0)} = \sigma_{\rm 8,0} \frac{X(z)}{X(z=0)} .
\label{eq:sigma8}
\end{equation}

Now we are in a position to check the effect of cosmic curvature on the growth of matter perturbations. To do so, we plot the comparison of $f$ between flat and non-flat $\Lambda$CDM models in Fig.~\ref{fig:f_cmp_Ok0}. In both left and right panels, we have fixed $H_0=70$ km$~$s$^{-1}~$Mpc$^{-1}$. The left and right panels correspond to fixed $\Omega_{\rm m0}=0.3$ and fixed $\Omega_{\rm Q0}=0.7$ respectively. The color combinations are the same as in previous figures. For both the cases (corresponding to left and right panels), the logarithmic growth rate is slower for positive $\Omega_{\rm k0}$ (i.e. for negative cosmic curvature) and faster for negative $\Omega_{\rm k0}$ (i.e. for positive cosmic curvature). Note that, this suppression or enhancement is almost independent of either we fix $\Omega_{\rm m0}$ or $\Omega_{\rm Q0}$. This was not the case for background expansion, we saw in Fig.~\ref{fig:H_cmp_Ok0} and Fig.~\ref{fig:DM_Ok0}. This makes the difference in effects of cosmic curvature on background and in growth. However, depending on whether we fix $\Omega_{\rm m0}$ or $\Omega_{\rm Q0}$, the amplitudes differ accordingly: the deviations are smaller in lower redshifts and slightly higher in higher redshifts for fixed $\Omega_{\rm m0}$; opposite happens for fixed $\Omega_{\rm Q0}$.

We also plot the comparison of $f\sigma_8$ between flat and non-flat $\Lambda$CDM models in Fig.~\ref{fig:fsigma8_cmp_Ok0}. For this plot, we have fixed $H_0=70$ km$~$s$^{-1}~$Mpc$^{-1}$ and $\sigma_{\rm 80}=0.8$. The left and right panels correspond to fixed $\Omega_{\rm m0}=0.3$ and fixed $\Omega_{\rm Q0}=0.7$ respectively. The color combinations are the same as in previous figures. The deviations are similar in behaviors as in $f$ in \autoref{fig:f_cmp_Ok0}, but the amplitudes are slightly different.

Note that, in \autoref{fig:H_cmp_Ok0}, \autoref{fig:DM_Ok0}, \autoref{fig:f_cmp_Ok0}, and \autoref{fig:fsigma8_cmp_Ok0}, we show deviations between zero and non-zero cosmic curvature by fixing either $\Omega_{\rm m0}$ or $\Omega_{\rm Q0}$ individually. It is not possible to fix both $\Omega_{\rm m0}$ and $\Omega_{\rm Q0}$ simultaneously while varying $\Omega_{\rm k0}$ due to the closure condition, mentioned in \autoref{eq:budget_constraint}. Therefore, the deviations shown in these plots do not represent the pure effect of cosmic curvature, but only provide an estimate of its possible impact. The best way to understand the effect of cosmic curvature is to test whether it is truly zero or non-zero by fitting all parameters freely to real cosmological data, as we do in the following sections.

To check the consistency of all the above equations, we compare our results with the CAMB\footnote{Here by CAMB we mean Code for Anisotropies in the Microwave Background; For details see \url{https://camb.readthedocs.io/en/latest/}} results in \autoref{sec-consistency_camb}.

\section{Observational data}
\label{sec-data}

In our analysis, we incorporate diverse observational datasets to robustly constrain the model parameters and cosmological nuisance parameters. We consider the supernova sample of Pantheon+ compilation, which compiles type Ia supernova observations of 1701 light curves for 1550 spectroscopically confirmed type Ia supernova (SNIa) in the redshift range $0.00122 \leq z \leq 2.26137$ \citep{Scolnic:2021amr}. These supernovae serve as standard candles, and their apparent peak absolute magnitudes $m_B$ vary with redshift \cite{Pan-STARRS1:2017jku}. The observed apparent magnitude $m_B$ is dependent on both the luminosity $d_L$ of the source at a given redshift and the nuisance parameter $M_B$, representing the peak absolute magnitude of SNIa given as
\begin{equation}
m_B = 5 \log_{10} \left( \frac{d_L}{{\rm Mpc}} \right) + 25 + M_B ,
\label{eq:snia}
\end{equation}
where $d_L$ is related to $D_M$ as
\begin{equation}
d_L = (1+z)D_M .
\label{eq:dL_to_DM}
\end{equation}
We additionally take into account the covariances existing between data points at varying redshifts to ensure a comprehensive analysis. The $m_B$ data and covariances (considering both statistical and systematic covariances) can be found at \url{https://github.com/PantheonPlusSH0ES/DataRelease}. We call this data as 'PantheonPlus'.

We consider data from cosmic chronometer (CC) observations with 32 Hubble parameter measurements spanning a diverse range of redshift values ($0.07 \leq z \leq 1.965$) \cite{Cao:2023eja}. Note that, within these 32 Hubble parameter measurements, 15 data points have correlations among themselves and we incorporate these into our analysis. We denote this dataset as 'CC'. The covariances can be found in \url{https://gitlab.com/mmoresco/CCcovariance/} \citep{Moresco:2020fbm,moresco2012improved,Moresco:2015cya,Moresco:2016mzx}.

We also consider baryon acoustic oscillations (BAO) data in our analysis. These observations provide insights into cosmological distances, specifically the angular diameter distance. The BAO dataset includes measurements in both the line of sight and transverse directions \cite{eBOSS:2020yzd}. The former is connected to the Hubble parameter, while the latter is linked to the angular diameter distance \cite{BOSS:2016wmc,eBOSS:2020yzd,Hou:2020rse}. For the BAO data, we consider Dark Energy Spectroscopic Instrument Data Release 2 (DESI DR2) data \citep{DESI:2025zgx}. The BAO observables are
\begin{eqnarray}
\tilde{D}_M &=& \frac{D_M}{r_d} ,
\label{eq:DMtilde} \\
\tilde{D}_H &=& \frac{D_H}{r_d} ,
\label{eq:DH_tilde} \\
& \text{with} & D_H = \frac{c}{H} , \\
\tilde{D}_V &=& \frac{D_V}{r_d} ,
\label{eq:DV_tilde} \\
&\text{with}& D_V = \left(z D_M^2 D_H\right)^{\frac{1}{3}} ,
\end{eqnarray}
where $r_d$ is the comoving sound horizon at the baryon drag epoch. Thus, BAO data are uncalibrated. To make these data calibrated, we consider constraints on $r_d$ parameter, obtained from Planck 2018 CMB analysis \citep{Planck:2018vyg} as
\begin{equation}
r_d = 147.43 \pm 0.25 ~ \text{[Mpc]} .
\label{eq:rd}
\end{equation}
This is the closest value used in DESI DR2 main paper \citep{DESI:2025zgx} for CMB $r_d$. Note that, we do not use CMB data itself, because our analysis is for late time only. We only use $r_d$ information because this is insensitive to late-time cosmology. We call this data as '$r_d$'. Also, we label these BAO observations for combined $\tilde{D}_M$, $\tilde{D}_H$, and $\tilde{D}_V$ as 'DESI DR2 BAO'.

We consider the logarithmic growth rate $f$ data for which we follow Avila et al. \citep{Avila:2022xad}. There are 11 uncorrelated $f$ data in the redshift range $0.013 \leq z \leq 1.4$.

Finally, we consider 20 uncorrelated $f\sigma_8$ data, spanning redshifts from $z=0.02$ to $z=1.944$. For this data, we also follow Avila et al. \citep{Avila:2022xad}.

\section{Results}
\label{sec-result}

\begin{figure*}
\centering
\includegraphics[width=0.98\textwidth]{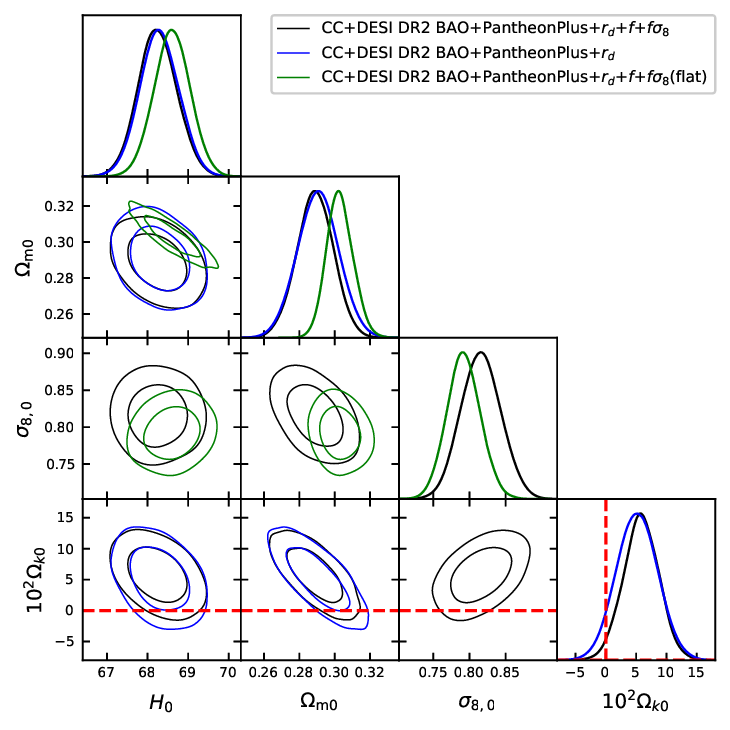}
\caption{
\label{fig:LCDM}
Constraints on the parameters of the $\Lambda$CDM model are depicted through various contour lines and marginalized probabilities through a triangle plot. The black, blue, and green lines correspond to the dataset combinations 'CC+DESI DR2 BAO+PantheonPlus+$r_d$+$f$+$f\sigma_{8}$' (considering cosmic curvature), 'CC+DESI DR2 BAO+PantheonPlus+$r_d$' (considering cosmic curvature), and 'CC+DESI DR2 BAO+PantheonPlus+$r_d$+$f$+$f\sigma_{8}$(flat)' (excluding cosmic curvature), respectively.
}
\end{figure*}

\begin{figure*}
\centering
\includegraphics[width=0.98\textwidth]{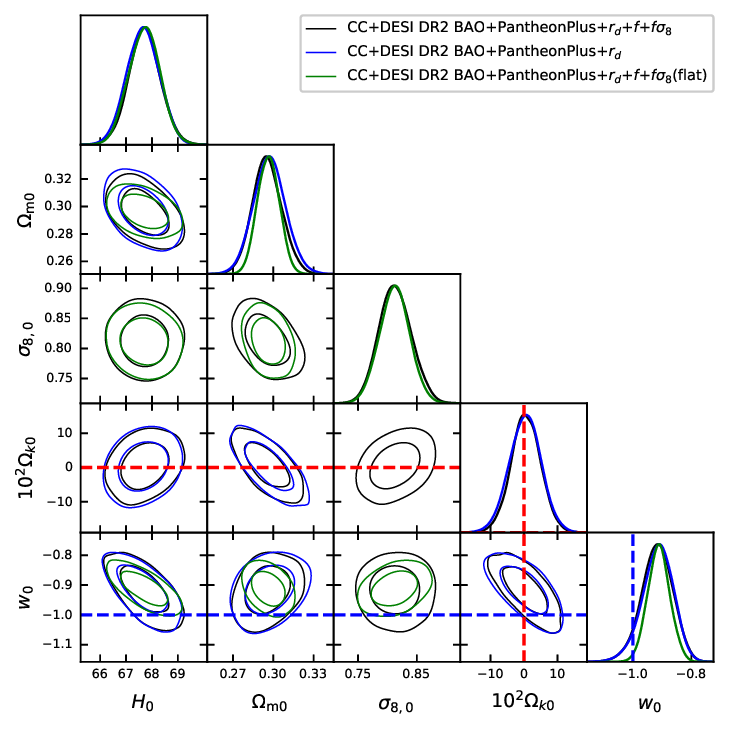}
\caption{
\label{fig:wCDM}
The constraints on the parameters of the $w_0$CDM model are illustrated through different contour lines and marginalized probabilities through a triangle plot. The black, blue, and green lines correspond to the dataset combinations 'CC+DESI DR2 BAO+PantheonPlus+$r_d$+$f$+$f\sigma_{8}$' (including cosmic curvature), 'CC+DESI DR2 BAO+PantheonPlus+$r_d$' (including cosmic curvature), and 'CC+DESI DR2 BAO+PantheonPlus+$r_d$+$f$+$f\sigma_{8}$(flat)' (excluding cosmic curvature), respectively.
}
\end{figure*}

\begin{figure*}
\centering
\includegraphics[width=0.98\textwidth]{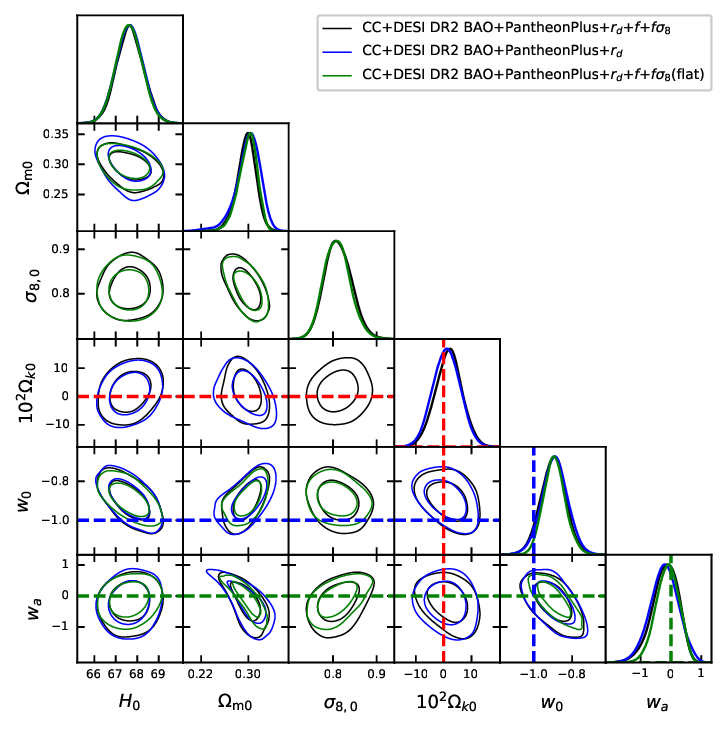}
\caption{
\label{fig:CPL}
The constraints on the parameters of the CPL model are depicted through various contour lines and marginalized probabilities through a triangle plot. The black, blue, and green lines correspond to the dataset combinations 'CC+DESI DR2 BAO+PantheonPlus+$r_d$+$f$+$f\sigma_{8}$' (considering cosmic curvature), 'CC+DESI DR2 BAO+PantheonPlus+$r_d$' (considering cosmic curvature), and 'CC+DESI DR2 BAO+PantheonPlus+$r_d$+$f$+$f\sigma_{8}$(flat)' (excluding cosmic curvature), respectively.
}
\end{figure*}

Using the above-mentioned observational data, we constrain all the model parameters in all three dark energy models. In Figure~\ref{fig:LCDM}, a triangle plot is presented to display the confidence contours of each pair of parameters and the marginal probability plots for each parameter in the $\Lambda$CDM model. The black lines represent the 'CC+DESI DR2 BAO+PantheonPlus+$r_d$+$f$+$f\sigma_{8}$' combination of the dataset, where cosmic curvature is taken into account in the analysis. The blue lines correspond to the 'CC+DESI DR2 BAO+PantheonPlus+$r_d$' combination of the dataset, incorporating cosmic curvature. The green lines depict the 'CC+DESI DR2 BAO+PantheonPlus+$r_d$+$f$+$f\sigma_{8}$(flat)' combination of the dataset, but in this case, cosmic curvature is not considered in the analysis (i.e., $\Omega_{\rm k0}=0$). The same color combinations are maintained for the same cases in the other two figures.

\begin{table*}
\begin{center}
\begin{tabular}{ |c|c|c|c|  }
\hline
Parameters & CC+DESI+PP+$r_d$+$f$+$f\sigma_{8}$ & CC+DESI+PP+$r_d$ & CC+DESI+PP+$r_d$+$f$+$f\sigma_{8}$(flat) \\
\hline
$H_0$ [km$~$s$^{-1}~$Mpc$^{-1}$] & $68.25\pm 0.48$ & $68.29\pm 0.48$ & $68.60\pm 0.45$ \\
\hline
$\Omega_{\rm m0}$ & $0.289\pm 0.010$ & $0.291\pm 0.012$ & $0.3029^{+0.0070}_{-0.0079}$ \\
\hline
$\sigma_{\rm 8,0}$ & $0.816\pm0.027$ & $-$ & $0.792\pm 0.023$ \\
\hline
$10^2\Omega_{\rm k0}$ & $5.8\pm 2.9$ & $5.1\pm3.4$ & $-$ \\
\hline
\end{tabular}
\end{center}
\caption{
1$\sigma$ bounds on the parameters of the $\Lambda$CDM model corresponding to three different combinations of datasets mentioned in the table.
}
\label{table:LCDM_all}
\end{table*}

We present all the marginalized 1$\sigma$ bounds on the parameters of the $\Lambda$CDM model in Table~\ref{table:LCDM_all}. Examining Figure~\ref{fig:LCDM} and Table~\ref{table:LCDM_all}, it is evident that the marginalized bounds on $H_0$ remain similar in all three cases. This suggests that the inclusion of $f+f\sigma_8$ data or the consideration of cosmic curvature in the analysis does not significantly alter the constraints $H_0$.

Note that, in \autoref{table:LCDM_all}, we denote DESI DR2 BAO shortly as 'DESI' and PantheonPlus shortly as 'PP'. We follow same notation in next two tables.

The constraints on $\Omega_{\rm k0}$ also exhibit similarity between the 'CC+DESI DR2 BAO+PantheonPlus+$r_d$+$f$+$f\sigma_{8}$' and 'CC+DESI DR2 BAO+PantheonPlus+$r_d$' combinations. Interestingly, we find around 2$\sigma$ and slightly more than 1.5$\sigma$ evidence for a non-zero cosmic curvature for the 'CC+DESI DR2 BAO+PantheonPlus+$r_d$+$f$+$f\sigma_{8}$' and 'CC+DESI DR2 BAO+PantheonPlus+$r_d$' combinations of data.

Notably, another interesting results emerge in the constraints on the $\Omega_{\rm m0}$ and $\sigma_{\rm 8,0}$ parameters. Constraints on $\Omega_{\rm m0}$ are similar for both 'CC+DESI DR2 BAO+PantheonPlus+$r_d$+$f$+$f\sigma_{8}$' and 'CC+DESI DR2 BAO+PantheonPlus+$r_d$' combinations but become notably tighter when cosmic curvature is excluded (and the mean value is slightly large), i.e., for 'CC+DESI DR2 BAO+PantheonPlus+$r_d$+$f$+$f\sigma_{8}$(flat)' (green line) combination. Similarly, the constraint on the $\sigma_{\rm 8,0}$ parameter is slightly tighter when the contribution of cosmic curvature is excluded. Furthermore, not only is the constraint tighter, but the mean value of $\sigma_{\rm 8,0}$ is higher when cosmic curvature is included in the analysis.

\begin{table*}
\begin{center}
\begin{tabular}{ |c|c|c|c|  }
\hline
Parameters & CC+DESI+PP+$r_d$+$f$+$f\sigma_{8}$ & CC+DESI+PP+$r_d$ & CC+DESI+PP+$r_d$+$f$+$f\sigma_{8}$(flat) \\
\hline
$H_0$ [km$~$s$^{-1}~$Mpc$^{-1}$] & $67.72\pm 0.61$ & $67.65\pm 0.63$ & $67.71\pm 0.61$ \\
\hline
$\Omega_{\rm m0}$ & $0.295\pm0.011$ & $0.297\pm0.012$ & $0.2963\pm0.0081$ \\
\hline
$\sigma_{\rm 8,0}$ & $0.813\pm0.028$ & $-$ & $0.812\pm0.026$ \\
\hline
$10^2 \Omega_{\rm k0}$ & $0.5\pm 4.4$ & $0.3\pm 4.8$ & $-$ \\
\hline
$w_0$ & $-0.918\pm0.054$ & $-0.916^{+0.058}_{-0.048}$ & $-0.912\pm 0.038$ \\
\hline
\end{tabular}
\end{center}
\caption{
1$\sigma$ bounds on the parameters of the $w_0$CDM model corresponding to three different combinations of datasets mentioned in the table.
}
\label{table:wCDM_all}
\end{table*}

In Figure~\ref{fig:wCDM}, we present triangle plots illustrating the $w_0$CDM model parameters for the same combinations of datasets, maintaining consistent color codes as in Figure~\ref{fig:LCDM}. Corresponding 1$\sigma$ bounds on the $w_0$CDM model parameters are tabulated in Table~\ref{table:wCDM_all}. Notably, an additional parameter, $w_0$, is introduced compared to the $\Lambda$CDM model. Similar to the $\Lambda$CDM model, the bounds on $H_0$ exhibit similarity across all three cases.

For $\Omega_{\rm m0}$ and $\sigma_{\rm 80}$ parameters, constraints are marginally tighter in the flat case compared to the non-flat cases, although not significantly so. The $\Lambda$CDM model, corresponding to $w_0=-1$, is almost 2$\sigma$ away from the mean values obtained in all three cases. Notably, the non-phantom behavior ($w_0>-1$) of dark energy appears more favorable compared to the phantom behavior ($w_0<-1$) in all three cases.

There is no evidence for non-zero cosmic curvature in $w_0$CDM model (while there is certain evidence for deviation from the $\Lambda$CDM model). This indirectly shows non-degeneracy between equation of state of dark energy and cosmic curvature. If these two were completely degenerate, then deviation from $w=-1$ would have impact on deviation from $\Omega_{\rm k0}=0$ and vice-versa. Clearly  this is not the case.

\begin{table*}
\begin{center}
\begin{tabular}{ |c|c|c|c|  }
\hline
Parameters & CC+DESI+PP+$r_d$+$f$+$f\sigma_{8}$ & CC+DESI+PP+$r_d$ & CC+DESI+PP+$r_d$+$f$+$f\sigma_{8}$(flat) \\
\hline
$H_0$ [km$~$s$^{-1}~$Mpc$^{-1}$] & $67.66\pm 0.62$ & $67.64\pm 0.63$ & $67.62\pm 0.61$ \\
\hline
$\Omega_{\rm m0}$ & $0.297^{+0.017}_{-0.014}$ & $0.299^{+0.023}_{-0.016}$ & $0.299^{+0.017}_{-0.013}$ \\
\hline
$\sigma_{\rm 8,0}$ & $0.812^{+0.029}_{-0.033}$ & $-$ & $0.809^{+0.028}_{-0.032}$ \\
\hline
$10^2 \Omega_{\rm k0}$ & $1.8\pm 4.8$ & $1.0\pm 4.9$ & $-$ \\
\hline
$w_0$ & $-0.899\pm0.067$ & $-0.891\pm0.069$ & $-0.892\pm 0.057$ \\
\hline
$w_a$ & $-0.20^{+0.48}_{-0.37}$ & $-0.23\pm0.46$ & $-0.12^{+0.40}_{-0.35}$ \\
\hline
\end{tabular}
\end{center}
\caption{
1$\sigma$ bounds on the parameters of the CPL model corresponding to three different combinations of datasets mentioned in the table.
}
\label{table:CPL_all}
\end{table*}

In Figure~\ref{fig:CPL}, we present triangle plots illustrating the CPL model parameters for the same combinations of datasets, maintaining consistent color codes as in Figures~\ref{fig:LCDM} and \ref{fig:wCDM}. Corresponding 1$\sigma$ bounds on the CPL model parameters are tabulated in Table \ref{table:CPL_all}. Similar to the $\Lambda$CDM and $w_0$CDM models, the bounds on $H_0$ remain similar across all three cases.

In contrast to the previous two models, the CPL model exhibits nearly identical bounds on the $\sigma_{\rm 8,0}$ parameter in both flat and non-flat cases. Moreover, for the $\Omega_{\rm m0}$ parameter, the CPL model displays a different behavior in constraints compared to the previous two models. The constraints on $\Omega_{\rm m0}$ are similar in both flat and non-flat cases, but excluding $f+f\sigma_8$ data results in looser constraints, with the mean value being comparatively lower.

Constraints on $w_0$ and $w_a$ parameters are similar across all three combinations. In the CPL model, we find that the $\Lambda$CDM model, representing $w_0=-1$ and $w_a=0$, is more than 1.5$\sigma$ away from the mean values obtained in all three cases. The result $w_0>-1$ and $w_a<0$ corresponds to an interesting result that equation of state of dark energy is non-phantom at lower redshift and phanotom at higher redshift.

Like $w_0$CDM model, there is no evidence for non-zero cosmic curvature in CPL model (while there is certain evidence for deviation from the $\Lambda$CDM model). This also indirectly shows non-degeneracy between equation of state of dark energy and cosmic curvature.

\section{Conclusion}
\label{sec-conclusion}

We delve into the impact of cosmic curvature on structure formation through a comprehensive analysis utilizing general relativistic first-order perturbation theory within the Newtonian gauge. Our computations involve continuity and Euler equations for a general fluid, as well as Einstein equations that incorporate cosmic curvature. Emphasizing the late-time dynamics of background expansion and first-order fluctuations, we scrutinize the evolution of matter density contrast in the presence of cosmic curvature. We explore this aspect with three distinct dark energy models $\Lambda$CDM, $w_0$CDM, and CPL.

We rewrite the evolution of matter density contrast considering cosmic curvature on the sub-Hubble scales using sub-Hubble assumptions. Employing proper initial conditions, we solve the evolution equation and conduct data analysis using key observational datasets: cosmic chronometers (CC), baryon acoustic oscillations from DESI DR2 BAO (including constraint on $r_d$), type Ia supernova observations from PantheonPlus, $f$ and $f\sigma_8$ data.

We find that constraints on $H_0$ remain largely unaffected by the inclusion of cosmic curvature or the presence of growth data ($f$ and $f\sigma_8$) across all three models considered in this analysis. However, in the case of $\Omega_{\rm m0}$ and $\sigma_{\rm 8,0}$ parameters, exclusion of cosmic curvature tightens constraints on $\Omega_{\rm m0}$ and $\sigma_{\rm 8,0}$ in $\Lambda$CDM and $w_0$CDM models compared to their non-flat counterparts. In contrast, for the CPL model, constraints on $\sigma_{\rm 8,0}$ remain consistent in all cases, while constraints on $\Omega_{\rm m0}$ are tighter, with a higher mean value when incorporating $f$ and $f\sigma_8$ data, regardless of cosmic curvature.

Interestingly, we find there is around 1.5$\sigma$ to 2$\sigma$ evidence of non-zero cosmic curvature in $\Lambda$CDM model.

Notably, in $w_0$CDM and CPL models, the $\Lambda$CDM model stands almost 2$\sigma$ and 1.5$\sigma$away respectively from the mean values obtained from data analysis for all data combinations. Moreover, in $w_0$CDM model, the non-phantom behavior of dark energy emerges as more favorable than the phantom behavior across all data combinations. In CPL model, equation of state of dark energy is non-phantom at lower redshift and phanotom at higher redshift.

Both in $w_0$CDM and CPL model, there is no evidence for non-zero cosmic curvature while there is certain evidence for dynamical dark energy equation of state of dark energy, which confirms the non-degeneracy between dynamical dark energy and cosmic curvature.

\section*{Acknowledgements}
The author would like to acknowledge IISER Kolkata for its financial support through the postdoctoral fellowship. The revision for this work was carried out during the author's tenure at UWC, partially supported by the South African Radio Astronomy Observatory and National Research Foundation (Grant No. 75415).

\appendix

\section{The metric to describe background and first-order scalar fluctuations}
\label{sec-metric}

We adopt the conformal Newtonian gauge to study the first-order general relativistic perturbations. Within this gauge, the line element, denoted as $ds$, characterizing solely scalar fluctuations, is expressed as

\begin{eqnarray}
ds^2 &=& \sigma \Bigg{[} -c^2(1+2\Phi)dt^2+a^2(1-2\Phi) \nonumber\\
&& \left( \frac{dr^2}{1-\kappa r^2}+r^2d\theta^2+r^2 \sin^2\theta d\phi^2 \right) \Bigg{]},
\label{eq:line_element}
\end{eqnarray}

\noindent
where $t$ represents cosmic time, $r$ denotes the comoving radial coordinate, and $\theta$ and $\phi$ stand for the comoving polar and azimuthal angles, respectively. $\kappa$ signifies the curvature of space-time, while $c$ is the speed of light in a vacuum and $a$ is the cosmic scale factor. The Bardeen potential is denoted as $\Phi$, and the parameter $\sigma$ is introduced to characterize the signature of the metric in the line element, defined as

\begin{equation}
    \sigma = \begin{cases}
    +1, & \mbox{for } (-,+,+,+) \mbox{ signature}, \\
    -1, & \mbox{for } (+,-,-,-) \mbox{ signature}. \\
    \end{cases}
    \label{eq:signature_sigma}
\end{equation}

\noindent
Note that in the line element given by Eq.~\eqref{eq:line_element}, we adopt the scalar-vector-tensor decomposition theory. This allows us to analyze scalar fluctuations independently. Additionally, our assumption of the absence of anisotropic stresses enables us to focus on a single degree of freedom in the scalar fluctuations, governed by the Bardeen potential $\Phi$. A background metric, representing the FLRW metric with curvature contributions, is characterized by $\Phi=0$.

The metric and its inverse can be expressed perturbatively as follows

\begin{eqnarray}
g_{\mu \nu} &=& \bar{g}_{\mu \nu} + \delta g_{\mu \nu} ,
\label{eq:metric_total} \\
g^{\mu \nu} &=& \bar{g}^{\mu \nu} + \delta g^{\mu \nu} ,
\label{eq:inverse_metric_total}
\end{eqnarray}

\noindent
respectively. We denote the background counterpart of a quantity with an over bar, and its first-order perturbation with the pre-factor $\delta$. Consistent notation is maintained throughout this study. The non-zero components of both the background metric and its inverse are listed below:

\begin{eqnarray}
\bar{g}_{t t} &=& -\sigma c^2, ~ \bar{g}_{r r} = \frac{\sigma a^2}{1-\kappa r^2}, ~ \bar{g}_{\theta \theta} = \sigma a^2 r^2, \nonumber\\
&& \bar{g}_{\phi \phi} = \sigma a^2 r^2 \sin^2\theta ,
\label{eq:bkg_metric} \\
\bar{g}^{t t} &=& - \frac{\sigma}{c^2}, ~ \bar{g}^{r r} = \frac{\sigma (1-\kappa r^2)}{a^2}, ~ \bar{g}^{\theta \theta} =  \frac{\sigma}{a^2 r^2}, \nonumber\\
&& \bar{g}^{\phi \phi} = \frac{\sigma}{a^2 r^2 \sin^2\theta} .
\label{eq:bkg_inverse_metric}
\end{eqnarray}

\noindent
The non-zero components of the first-order metric and its inverse are listed below:

\begin{eqnarray}
\delta g_{t t} &=& -2 \sigma c^2 \Phi, ~ \delta g_{r r} = - \frac{2 \sigma a^2 \Phi}{1-\kappa r^2}, ~ \delta g_{\theta \theta} = -2 \sigma a^2 r^2 \Phi, \nonumber\\
&& \delta g_{\phi \phi} = -2 \sigma a^2 r^2 \left( \sin^2\theta \right) \Phi ,
\label{eq:pert_metric} \\
\delta g^{t t} &=& \frac{2 \sigma \Phi}{c^2}, ~ \delta g^{r r} = \frac{2 \sigma (1-\kappa r^2) \Phi}{a^2}, ~ \delta g^{\theta \theta} =  \frac{2 \sigma \Phi}{a^2 r^2}, \nonumber\\
&& \delta g^{\phi \phi} = \frac{2 \sigma \Phi}{a^2 r^2 \sin^2\theta} .
\label{eq:pert_inverse_metric}
\end{eqnarray}

\section{The background and the first order Einstein tensor}
\label{sec-Einstein_tensor}

With the components of the metric and its inverse in hand, we proceed to compute the components of the Einstein tensor perturbatively, denoted as

\begin{equation}
G^{\mu}_{\nu} = \bar{G}^{\mu}_{\nu} + \delta G^{\mu}_{\nu} .
\label{eq:EN_tensor}
\end{equation}

\noindent
The non-zero components of the Einstein tensor are provided below:

\begin{eqnarray}
\bar{G}^t_t &=& -3 \sigma \left( \frac{H^2}{c^2} + \frac{\kappa}{a^2} \right),
\label{eq:Bkg_Einstein_tensor_1} \\
\bar{G}^r_r &=& \bar{G}^{\theta}_{\theta} = \bar{G}^{\phi}_{\phi} = - \sigma \left( \frac{2\dot{H}+3H^2}{c^2} + \frac{\kappa}{a^2} \right),
\label{eq:Bkg_Einstein_tensor_2}
\end{eqnarray}

\noindent
where an over dot on a quantity indicates the partial derivative of that quantity w.r.t cosmic time $t$. It's important to note that for background quantities, the partial derivative w.r.t $t$ is equivalent to the total derivative w.r.t $t$ because all background quantities are functions of cosmic time $t$. On the other hand, perturbed quantities are functions of all four coordinates—$t$, $r$, $\theta$, and $\phi$. Here, $H$ represents the Hubble parameter, defined as $\frac{\dot{a}}{a}$. The $t$-$t$ component of the first-order Einstein tensor is expressed as

\begin{equation}
\delta G^t_t = 2 \sigma \left[ \frac{3H \left( \frac{\partial \Phi}{\partial t}+H\Phi \right) }{c^2} - \frac{\vec{\nabla}^2\Phi+3\kappa \Phi}{a^2} \right] ,
\label{eq:pert_EN_tensor_tt}
\end{equation}

\noindent
where $\vec{\nabla}^2f$ represents the comoving gradient square, corresponding to the three-dimensional space, of a quantity $f$, and is defined as

\begin{eqnarray}
\vec{\nabla}^2 f &=& \frac{1}{h_r h_{\theta} h_{\phi}} \Bigg{[} \frac{\partial}{\partial r} \left( \frac{h_{\theta} h_{\phi}}{h_r} \frac{\partial f}{\partial r} \right) + \frac{\partial}{\partial \theta} \left( \frac{h_{\phi} h_r}{h_{\theta}} \frac{\partial f}{\partial \theta} \right) \nonumber\\
&& + \frac{\partial}{\partial \phi} \left( \frac{h_r h_{\theta}}{h_{\phi}} \frac{\partial f}{\partial \phi} \right) \Bigg{]} , \nonumber\\
&=& \frac{1}{r^2} \Bigg{[} r^2(1-\kappa r^2) \frac{\partial^2 f}{\partial r^2} + r(2-3\kappa r^2) \frac{\partial f}{\partial r} + \frac{\partial^2 f}{\partial \theta^2} \nonumber\\
&& + \left( \cot\theta \right) \frac{\partial f}{\partial \theta} + \frac{ \frac{\partial^2 f}{\partial \phi^2} }{\sin^2\theta} \Bigg{]} .
\label{eq:grad_sqr_f_curvi}
\end{eqnarray}

\noindent
Here, the factors of the curvilinear coordinates are given as

\begin{eqnarray}
h_r &=& \frac{1}{\sqrt{1-\kappa r^2}} , \nonumber\\
h_{\theta} &=&  r, \nonumber\\
h_{\phi} &=& r \sin\theta .
\label{eq:hr_htheta_hphi}
\end{eqnarray}

\noindent
The $r$-$t$ and $t$-$r$ components of the first-order Einstein tensor are given as

\begin{eqnarray}
\delta G^r_t &=& \frac{2\sigma(1-\kappa r^2)}{a^2} \frac{\partial}{\partial r} \left( \frac{\partial \Phi}{\partial t}+H\Phi \right) \nonumber\\
&=& - \frac{c^2(1-\kappa r^2)}{a^2} \delta G^t_r .
\label{eq:pert_EN_tensor_rt_tr}
\end{eqnarray}

\noindent
The $\theta$-$t$ and $t$-$\theta$ components of the first-order Einstein tensor are given as

\begin{equation}
\delta G^{\theta}_t = \frac{2\sigma}{a^2r^2} \frac{\partial}{\partial \theta} \left( \frac{\partial \Phi}{\partial t}+H\Phi \right) = - \frac{c^2}{a^2r^2} \delta G^t_{\theta} .
\label{eq:pert_EN_tensor_thetat_ttheta}
\end{equation}

\noindent
The $\phi$-$t$ and $t$-$\phi$ components of the first-order Einstein tensor are given as

\begin{eqnarray}
\delta G^{\phi}_t &=& \frac{2\sigma}{a^2r^2 \sin^2\theta} \frac{\partial}{\partial \phi} \left( \frac{\partial \Phi}{\partial t}+H\Phi \right) \nonumber\\
&=& - \frac{c^2}{a^2r^2 \sin^2\theta} \delta G^t_{\phi} .
\label{eq:pert_EN_tensor_phit_tphi}
\end{eqnarray}

\noindent
The $r$-$r$, $\theta$-$\theta$, and $\phi$-$\phi$ components of the first-order Einstein tensor are given as

\begin{align}
\label{eq:pert_EN_tensor_diag}
\delta G^r_r &= \delta G^{\theta}_{\theta} = \delta G^{\phi}_{\phi} \nonumber\\
&= 2 \sigma \left[ \frac{ \frac{\partial^2\Phi}{\partial t^2} + 4 H \frac{\partial \Phi}{\partial t} + (2\dot{H}+3H^2) \Phi }{c^2} - \frac{\kappa \Phi}{a^2} \right] ,
\end{align}

\noindent
All other components of the first-order Einstein tensor are trivially zero.

\section{The description of a perfect fluid}
\label{sec-all_perfect_fluid}

In this section, we elucidate the dynamics of a perfect fluid within this metric framework.

\subsection{The background and the first order velocity field of a perfect fluid}
\label{sec-velocity_perfect_fluid}

To characterize a perfect fluid, let's express its four-velocity components, both covariant and contravariant, perturbatively as

\begin{eqnarray}
U^{\mu} &=& \bar{U}^{\mu} + \delta U^{\mu} ,
\label{eq:velocity_defn_upper} \\
U_{\mu} &=& \bar{U}_{\mu} + \delta U_{\mu} .
\label{eq:velocity_defn_lower}
\end{eqnarray}

\noindent
The spatial components of the contravariant four-velocity become zero when studying only scalar fluctuations. Thus, we have

\begin{equation}
\bar{U}^r = \bar{U}^{\theta} = \bar{U}^{\phi} = 0 .
\label{eq:velocity_bkg_upper_spatial}
\end{equation}

\noindent
The time component of the contravariant four-velocity can be computed using the identity provided as

\begin{equation}
\bar{g}_{\mu \nu} \bar{U}^{\mu} \bar{U}^{\nu} = -\sigma c^2 .
\label{eq:velocity_bkg_constraint}
\end{equation}

\noindent
Applying this identity, we derive an algebraic equation for $\bar{U}^t$ as

\begin{equation}
\left( \bar{U}^t \right)^2 = 1 ,
\label{eq:velocity_bkg_upper_t_cndn}
\end{equation}

\noindent
and we opt for the positive solution of the above equation, yielding

\begin{equation}
\bar{U}^t = 1 .
\label{eq:velocity_bkg_upper_t}
\end{equation}

\noindent
Utilizing the contruction with the background metric, i.e., $\bar{U}_{\mu}=\bar{g}_{\mu \nu} \bar{U}^{\nu}$, we obtain all components of the covariant four-velocity as

\begin{equation}
\bar{U}_t = -\sigma c^2, \hspace{1 cm} \bar{U}_r = \bar{U}_{\theta} = \bar{U}_{\phi} = 0 .
\label{eq:velocity_bkg_lower}
\end{equation}

Similarly, the first-order time component of the contravariant four-vector is computed using the provided identity

\begin{equation}
g_{\mu \nu} U^{\mu} U^{\nu} = -\sigma c^2 .
\label{eq:velocity_total_constraint}
\end{equation}

\noindent
And we have

\begin{equation}
\delta U^t = -\Phi ,
\label{eq:velocity_pert_upper_t}
\end{equation}

\noindent
By using $U_{\mu}=g_{\mu \nu} U^{\nu}$, we get

\begin{eqnarray}
\delta U_t &=& -\sigma c^2 \Phi, 
\label{eq:velocity_prt_lower_t} \\
\delta U_r &=& \frac{\sigma a^2 \delta U^r}{1-\kappa r^2}, \hspace{0.2 cm} \delta U_{\theta} = \sigma a^2 r^2 \delta U^{\theta}, \nonumber\\
&& \delta U_{\phi} = \sigma a^2 r^2 \left( \sin^2\theta \right) \delta U^{\phi} .
\label{eq:velocity_prt_lower_others}
\end{eqnarray}

\noindent
The curl-free part of the velocity contributes to the scalar fluctuations. Therefore, we opt for

\begin{eqnarray}
&& \delta U_r = - \sigma \frac{\partial (a v)}{\partial r}, \hspace{0.5 cm} \delta U_{\theta} = - \sigma \frac{\partial (a v)}{\partial \theta}, \nonumber\\
&& \delta U_{\phi} = - \sigma \frac{\partial (a v)}{\partial \phi} .
\label{eq:velocity_prt_lower_redefine}
\end{eqnarray}

\noindent
Now, employing the relations in Eq.~\eqref{eq:velocity_prt_lower_others}, we determine the contravariant spatial components of the four-velocity as

\begin{eqnarray}
&& \delta U^r = - \frac{1-\kappa r^2}{a} \frac{\partial v}{\partial r}, \hspace{0.5 cm} \delta U^{\theta} = - \frac{1}{a r^2} \frac{\partial v}{\partial \theta}, \nonumber\\
&& \delta U^{\phi} = - \frac{1}{a r^2 \sin^2\theta} \frac{\partial v}{\partial \phi} .
\label{eq:velocity_prt_upper_final}
\end{eqnarray}

\subsection{The energy-momentum tensor of a perfect fluid}
\label{sec-EM_tensor_perfect_fluid}

The energy-momentum tensor of a perfect fluid can be expressed in the following form

\begin{equation}
T^{\mu}_{\nu} = \left( \rho + \frac{P}{c^2} \right) U^{\mu} U_{\nu} + \sigma P \delta^{\mu}_{\nu} ,
\label{eq:EM_tensor_defn}
\end{equation}

\noindent
where $\rho$ and $P$ represent the volumetric mass density and pressure of the perfect fluid, respectively, while $\delta$ denotes the usual Kronecker delta symbol. Note that, this $\delta$ is different from the previously used $\delta$ in front of any quantity. The volumetric mass density and pressure of the perfect fluid are expressed in perturbative order as

\begin{eqnarray}
\rho &=& \bar{\rho} + \delta \rho ,
\label{eq:rho_tot} \\
P &=& \bar{P} + \delta P ,
\label{eq:P_tot}
\end{eqnarray}

\noindent
respectively. Similarly, the energy-momentum tensor of the perfect fluid can be expressed in perturbative order as

\begin{equation}
T^{\mu}_{\nu} = \bar{T}^{\mu}_{\nu} + \delta T^{\mu}_{\nu} ,
\label{eq:E_M_tensor_total}
\end{equation}

\noindent
Utilizing Eqs.~\eqref{eq:velocity_bkg_upper_spatial}, ~\eqref{eq:velocity_bkg_upper_t}, ~\eqref{eq:velocity_bkg_lower}, ~\eqref{eq:velocity_pert_upper_t}, ~\eqref{eq:velocity_prt_lower_t}, ~\eqref{eq:velocity_prt_upper_final}, ~\eqref{eq:rho_tot}, and \eqref{eq:P_tot} in Eq.\eqref{eq:EM_tensor_defn}, we obtain components of the energy-momentum tensor for both the background and first-order counterpart. The non-zero background components of the energy-momentum tensor for the perfect fluid are listed below

\begin{eqnarray}
\bar{T}^t_t &=& - \sigma c^2 \bar{\rho} ,
\label{eq:Bkg_E_M_tensor_1} \\
\bar{T}^r_r &=& \bar{T}^{\theta}_{\theta} = \bar{T}^{\phi}_{\phi} = \sigma \bar{P} .
\label{eq:Bkg_E_M_tensor_2}
\end{eqnarray}

\noindent
The $t$-$t$ component of the first-order energy-momentum tensor for the perfect fluid is given below

\begin{equation}
\delta T^t_t = - \sigma c^2 \delta \rho .
\label{eq:pert_E_M_tt}
\end{equation}

\noindent
The $r$-$t$ and $t$-$r$ components of the first-order energy-momentum tensor for the perfect fluid are given below

\begin{equation}
\delta T^r_t = \frac{ \sigma (1-\kappa r^2) (c^2 \bar{\rho} +\bar{P}) \frac{\partial v}{\partial r} }{a} = - \frac{c^2(1-\kappa r^2)}{a^2} \delta T^t_r .
\label{eq:pert_E_M_rt_tr}
\end{equation}

\noindent
The $\theta$-$t$ and $t$-$\theta$ components of the first-order energy-momentum tensor for the perfect fluid are given below

\begin{equation}
\delta T^{\theta}_t = \frac{ \sigma (c^2 \bar{\rho} +\bar{P}) \frac{\partial v}{\partial \theta} }{a r^2} = - \frac{c^2}{a^2r^2} \delta T^t_{\theta} .
\label{eq:pert_E_M_thetat_ttheta}
\end{equation}

\noindent
The $\phi$-$t$ and $t$-$\phi$ components of the first-order energy-momentum tensor for the perfect fluid are given below

\begin{equation}
\delta T^{\phi}_t = \frac{ \sigma (c^2 \bar{\rho} +\bar{P}) \frac{\partial v}{\partial \phi} }{a r^2 \sin^2\theta} = - \frac{c^2}{a^2r^2 \sin^2\theta} \delta T^t_{\phi} ,
\label{eq:pert_E_M_phit_tphi}
\end{equation}

\noindent
The $r$-$r$, $\theta$-$\theta$, and $\phi$-$\phi$ components of the first-order energy-momentum tensor for the perfect fluid are given below

\begin{equation}
\delta T^r_r = \delta T^{\theta}_{\theta} = \delta T^{\phi}_{\phi} = \sigma \delta P .
\label{eq:pert_E_M_diag}
\end{equation}

\noindent
All other components of the first-order energy-momentum tensor for the perfect fluid are trivially zero.

\subsection{The continuity equation for a perfect fluid}
\label{sec-continuity_perfect_fluid}

In this study, we assume that there is no interaction of the perfect fluid with other fields or fluids present in the universe. Therefore, the divergence of the energy-momentum tensor for the perfect fluid is zero, i.e.,

\begin{equation}
\nabla _{\mu} T^{\mu}_{\nu} = 0 .
\label{eq:divergence_free_E_M}
\end{equation}

\noindent
Analyzing the above equation order by order, we arrive at the background condition

\begin{equation}
\bar{\nabla} _{\mu} \bar{T}^{\mu}_{\nu} = 0 .
\label{eq:cntn_defn_bkg}
\end{equation}

\noindent
The above condition corresponds to four equations. The $t$ part of this equation corresponds to

\begin{equation}
\dot{\bar{\rho}} + 3 H \left( \bar{\rho} +\frac{ \bar{P} }{ c^2 } \right) = 0 .
\label{eq:cntn_bkg}
\end{equation}

\noindent
This constitutes the continuity equation for the perfect fluid at the background level. Now, we establish a relationship between the background pressure and energy density of the perfect fluid using a quantity called the equation of state, denoted by $w$ and defined as

\begin{equation}
\bar{P} = w \bar{\rho} c^2 .
\label{eq:defn_EoS}
\end{equation}

\noindent
Utilizing Eq.\eqref{eq:defn_EoS}, Eq.\eqref{eq:cntn_bkg} can be rewritten as

\begin{equation}
\dot{\bar{\rho}} + 3 H (1+w) \bar{\rho} = 0 .
\label{eq:cntn_bkg_2}
\end{equation}

\noindent
Similarly, we obtain the first-order continuity equation for the perfect fluid as

\begin{equation}
\dot{(\delta \rho)} + 3 H \left( \delta \rho +\frac{ \delta P }{ c^2 } \right) - \frac{(1+w)\bar{\rho} (\theta_g + 3 a \dot{\Phi})}{a} = 0 ,
\label{eq:cntn_pert_pre}
\end{equation}

\noindent
where

\begin{eqnarray}
\theta_g = \vec{\nabla}^2 v &=& \frac{1}{r^2} \Bigg{[} r^2(1-\kappa r^2) \frac{\partial^2 v}{\partial r^2} + r(2-3\kappa r^2) \frac{\partial v}{\partial r} \nonumber\\
&& + \frac{\partial^2 v}{\partial \theta^2} + \left( \cot\theta \right) \frac{\partial v}{\partial \theta} + \frac{ \frac{\partial^2 v}{\partial \phi^2} }{\sin^2\theta} \Bigg{]} .
\label{eq:defn_theta_g}
\end{eqnarray}

\noindent
Now, we define the density contrast, $\delta_g$, for the perfect fluid as

\begin{equation}
\delta \rho = \bar{\rho} \delta_g .
\label{eq:defn_delta_rho}
\end{equation}

\noindent
Applying the above definition, Eq.\eqref{eq:cntn_pert_pre} can be rewritten as

\begin{equation}
\dot{\delta}_g - 3 w H \delta_g - \frac{(1+w)\theta_g}{a} -3(1+w)\dot{\Phi} + \frac{3 H (\delta P)}{c^2 \bar{\rho}} = 0 .
\label{eq:cntn_pert}
\end{equation}

\subsection{The Euler equations for a perfect fluid}
\label{sec-Euler_equations_perfect_fluid}

The spatial parts of Eq.\eqref{eq:divergence_free_E_M} are referred to as the Euler equations for the perfect fluid. The background Euler equations for the perfect fluid are trivially zero, i.e.,

\begin{equation}
\bar{\nabla} _{\mu} \bar{T}^{\mu}_{r} = \bar{\nabla} _{\mu} \bar{T}^{\mu}_{\theta} = \bar{\nabla} _{\mu} \bar{T}^{\mu}_{\phi} = 0 .
\label{eq:Euler_eqns_cndn_bkg}
\end{equation}

\noindent
So, we do not obtain any corresponding equations. The first-order Euler equations corresponding to the conditions $\nabla _{\mu} T^{\mu}_{r} = 0$, $\nabla _{\mu} T^{\mu}_{\theta} = 0$, and $\nabla _{\mu} T^{\mu}_{\phi} = 0$ are given as

\begin{eqnarray}
\frac{\partial A}{\partial r} &=& 0 ,
\label{eq:Euler_eqn_pert_r} \\
\frac{\partial A}{\partial \theta} &=& 0 ,
\label{eq:Euler_eqn_pert_theta} \\
\frac{\partial A}{\partial \phi} &=& 0 ,
\label{eq:Euler_eqn_pert_phi}
\end{eqnarray}

\noindent
respectively, where we have

\begin{equation}
A = \dot{v} + \left[ (1-3w)H + \frac{\dot{w}}{1+w} \right] v - \frac{c^2 \Phi}{a} - \frac{\delta P}{a(1+w)\bar{\rho}} .
\label{eq:expression_A}
\end{equation}

\noindent
As all three spatial first derivatives of $A$ are individually zero, we can further manipulate these equations into the following form to obtain a single useful Euler equation given as

\begin{eqnarray}
&& \frac{1}{h_r h_{\theta} h_{\phi}} \Bigg{[} \frac{\partial}{\partial r} \left( \frac{h_{\theta} h_{\phi}}{h_r} \frac{\partial A}{\partial r} \right) + \frac{\partial}{\partial \theta} \left( \frac{h_{\phi} h_r}{h_{\theta}} \frac{\partial A}{\partial \theta} \right) \nonumber\\
&& + \frac{\partial}{\partial \phi} \left( \frac{h_r h_{\theta}}{h_{\phi}} \frac{\partial A}{\partial \phi} \right) \Bigg{]} = \vec{\nabla}^2 A = 0 .
\label{eq:Euler_manipulations}
\end{eqnarray}

\noindent
Finally, by substituting the expression of $A$ from Eq.\eqref{eq:expression_A} into Eq.\eqref{eq:Euler_manipulations}, we obtain a simplified form of the Euler equation for the perfect fluid given as

\begin{eqnarray}
&& \dot{\theta}_g + \left[ (1-3w)H + \frac{\dot{w}}{1+w} \right] \theta_g - \frac{c^2 \vec{\nabla}^2 \Phi}{a} \nonumber\\
&& - \frac{ \vec{\nabla}^2(\delta P)}{a(1+w)\bar{\rho}} =0 .
\label{eq:final_Euler_equation}
\end{eqnarray}

\section{Matter as a perfect fluid}
\label{sec-matter}

The matter components of the Universe can be modeled as a perfect fluid. In this context, we assume the matter is cold, implying zero background Pressure, resulting in an equation of state $w=0$, and also, first-order pressure is zero for the matter. Additionally, we assume that matter does not interact with other fields. Because matter is considered a perfect fluid, all the equations describing a perfect fluid hold for the matter, with the additional constraints $w=0$, $\bar{P}=0$, and $\delta P=0$. Moreover, $\dot{w}=0$ and $\vec{\nabla}^2(\delta P)=0$. Throughout this study, we denote all non-zero quantities related to matter with the subscript "m". Using these considerations, let's outline the key equations for the matter counterpart. By substituting $w=0$ into Eq.\eqref{eq:cntn_bkg_2}, we obtain the background continuity equation for matter given as

\begin{equation}
\dot{\bar{\rho}}_m + 3 H \bar{\rho}_m = 0 .
\label{eq:cntn_bkg_matter}
\end{equation}

\noindent
Similarly, substituting $w=0$ and $\delta P=0$ into Eq.\eqref{eq:cntn_pert}, we derive the first-order continuity equation for matter given as

\begin{equation}
\dot{\delta}_m - \frac{\theta_m}{a} -3\dot{\Phi} = 0 .
\label{eq:cntn_pert_matter}
\end{equation}

\noindent
Similarly, substituting $w=0$, $\dot{w}=0$, and $\vec{\nabla}^2(\delta P)=0$ into Eq.\eqref{eq:final_Euler_equation}, we derive the first-order Euler equation for matter given as

\begin{equation}
\dot{\theta}_m + H \theta_m - \frac{c^2 \vec{\nabla}^2 \Phi}{a} = 0 .
\label{eq:Euler_equation_matter}
\end{equation}

Let's further manipulate Eqs.\eqref{eq:cntn_pert_matter} and \eqref{eq:Euler_equation_matter} to derive a more useful differential equation for the matter. To do this, we first take the time derivative of Eq.\eqref{eq:cntn_pert_matter}, resulting in

\begin{equation}
\ddot{\delta}_m - \frac{\dot{\theta}_m}{a} + \frac{H\theta_m}{a} -3 \ddot{\Phi} = 0 .
\label{eq:dm_double_dot_1}
\end{equation}

\noindent
By utilizing Eq.\eqref{eq:Euler_equation_matter}, we obtain

\begin{equation}
\dot{\theta}_m = \frac{c^2 \vec{\nabla}^2 \Phi}{a} - H\theta_m .
\label{eq:sol_thetam_dot}
\end{equation}

\noindent
Substituting the above equation into Eq.\eqref{eq:dm_double_dot_1}, we obtain

\begin{equation}
\ddot{\delta}_m + \frac{2 H\theta_m}{a} - \frac{c^2 \vec{\nabla}^2 \Phi}{a^2} -3 \ddot{\Phi} = 0 .
\label{eq:dm_double_dot_2}
\end{equation}

\noindent
By employing Eq.\eqref{eq:cntn_pert_matter}, we obtain

\begin{equation}
\theta_m = a(\dot{\delta}_m-3\dot{\Phi}) .
\label{eq:sol_thetam}
\end{equation}

\noindent
Substituting the above equation into Eq.\eqref{eq:dm_double_dot_2}, we obtain a second-order differential equation for $\delta_m$ that is independent of $\theta_m$, given as

\begin{equation}
\ddot{\delta}_m + 2H \dot{\delta}_m - \frac{c^2 \vec{\nabla}^2 \Phi}{a^2} -6 H \dot{\Phi} -3 \ddot{\Phi} = 0 .
\label{eq:dm_double_dot_Final}
\end{equation}

\section{Late-time dynamics: Einstein equations}
\label{sec-late_time_Einstein_equations}

We are now at a stage to derive the Einstein equations. This study emphasizes the dynamics of the late-time Universe, where we can neglect the contribution of radiation. We exclusively consider matter and dark energy, where 'matter' refers to cold dark matter and baryons combined. Henceforth, we use the abbreviation 'DE' for dark energy and denote any quantity related to dark energy with the subscript 'Q'. In this scenario, the Einstein equations are obtained perturbatively as follows

\begin{eqnarray}
G^{\mu}_{\nu} &=& \frac{8\pi G}{c^4} \left[ T^{\mu}_{\nu}(\rm matter) + T^{\mu}_{\nu}(\rm DE) \right],
\label{eq:Einstein_eqns_defn_tot} \\
\bar{G}^{\mu}_{\nu} &=& \frac{8\pi G}{c^4} \left[ \bar{T}^{\mu}_{\nu}(\rm matter) + \bar{T}^{\mu}_{\nu}(\rm DE) \right],
\label{eq:Einstein_eqns_defn_bkg} \\
\delta G^{\mu}_{\nu} &=& \frac{8\pi G}{c^4} \left[ \delta T^{\mu}_{\nu}(\rm matter) + \delta T^{\mu}_{\nu}(\rm DE) \right],
\label{eq:Einstein_eqns_defn_pert}
\end{eqnarray}

\noindent
where $G$ represents Newton's gravitational constant. Utilizing all the above equations, we finally obtain the two independent non-trivial Einstein equations for the background evolution, given below

\begin{eqnarray}
3H^2 &=& 8\pi G (\bar{\rho}_m+\bar{\rho}_Q) - \frac{3c^2 \kappa}{a^2} ,
\label{eq:Friedmann_1} \\
2\dot{H} + 3H^2 &=& -\frac{8\pi G}{c^2} \bar{P}_Q - \frac{c^2 \kappa}{a^2} .
\label{eq:Friedmann_2}
\end{eqnarray}

\noindent
The $t$-$t$ component of the first-order perturbation in the Einstein equations is given as

\begin{eqnarray}
&& \frac{c^2\left(\vec{\nabla}^2 \Phi+3\kappa \Phi\right)}{a^2} -3H(\dot{\Phi}+H\Phi) = 4\pi G (\delta \rho_m + \delta \rho_Q) \nonumber\\
&& = 4\pi G (\bar{\rho}_m \delta_m + \bar{\rho}_Q \delta_Q) .
\label{eq:pert_Einstein_eqn_1}
\end{eqnarray}

\noindent
The three independent equations corresponding to the $r$-$t$ (or $t$-$r$), $\theta$-$t$ (or $t$-$\theta$), and $\phi$-$t$ (or $t$-$\phi$) components of the first-order perturbation in the Einstein equations are given as

\begin{eqnarray}
\frac{\partial B}{\partial r} &=& 0 ,
\label{eq:En_eqn_pert_r} \\
\frac{\partial B}{\partial \theta} &=& 0 ,
\label{eq:En_eqn_pert_theta} \\
\frac{\partial B}{\partial \phi} &=& 0 ,
\label{eq:En_eqn_pert_phi}
\end{eqnarray}

\noindent
respectively, where we have

\begin{equation}
B = \dot{\Phi} +  H\Phi - \frac{4\pi G a}{c^2} \left[ \bar{\rho}_m v_m + \bar{\rho}_Q (1+w_Q) v_Q \right] .
\label{eq:expression_B}
\end{equation}

\noindent
Eqs.\eqref{eq:En_eqn_pert_r}, \eqref{eq:En_eqn_pert_theta}, and \eqref{eq:En_eqn_pert_phi} can be manipulated to be written in a single equation, as

\begin{eqnarray}
&& \frac{1}{h_r h_{\theta} h_{\phi}} \Bigg{[} \frac{\partial}{\partial r} \left( \frac{h_{\theta} h_{\phi}}{h_r} \frac{\partial B}{\partial r} \right) + \frac{\partial}{\partial \theta} \left( \frac{h_{\phi} h_r}{h_{\theta}} \frac{\partial B}{\partial \theta} \right) \nonumber\\
&& + \frac{\partial}{\partial \phi} \left( \frac{h_r h_{\theta}}{h_{\phi}} \frac{\partial B}{\partial \phi} \right) \Bigg{]} = \vec{\nabla}^2 B = 0 .
\label{eq:En_Eqn_2_manipulations}
\end{eqnarray}

\noindent
Substituting the expression of $B$ from Eq.\eqref{eq:expression_B} into Eq.\eqref{eq:En_Eqn_2_manipulations}, we finally obtain a simplified form of all three Einstein equations in a useful single equation, as

\begin{equation}
c^2 \vec{\nabla}^2 (\dot{\Phi}+H\Phi) = 4\pi G a \left[ \bar{\rho}_m \theta_m + \bar{\rho}_Q (1+w_Q) \theta_Q \right] .
\label{eq:pert_Einstein_eqn_2}
\end{equation}
\\
\noindent
Finally, a single independent equation from either the $r$-$r$, $\theta$-$\theta$, or $\phi$-$\phi$ components of the first-order perturbation in the Einstein equations is given as

\begin{equation}
\ddot{\Phi} + 4H\dot{\Phi} + (2 \dot{H} +3H^2) \Phi - \frac{c^2\kappa \Phi}{a^2} = \frac{4\pi G}{c^2} \delta P_Q .
\label{eq:pert_Einstein_eqn_3}
\end{equation}

\noindent
All other Einstein equations are trivially zero, so no additional equations arise.

\section{Late-time dynamics: sub-Hubble approximation}
\label{sec-late_time_sub-Hubble}

We are now focusing on sub-Hubble scales to study the evolution of perturbations, particularly matter density contrast. On sub-Hubble scales, we assume that the spatial derivatives of a quantity are significantly larger than its time derivative. Therefore, we can neglect the terms $H \dot{\Phi}$ and $\ddot{\Phi}$ compared to the term $\frac{c^2 \vec{\nabla}^2 \Phi}{a^2}$ in Eq.\eqref{eq:dm_double_dot_Final}. Thus, in the sub-Hubble limit, Eq.\eqref{eq:dm_double_dot_Final} can be approximated as

\begin{equation}
\ddot{\delta}_m + 2H \dot{\delta}_m - \frac{c^2 \vec{\nabla}^2 \Phi}{a^2} \simeq 0 .
\label{eq:dm_double_dot_sub_Hubble_1}
\end{equation}

\noindent
Applying similar approximations, in the sub-Hubble limit, Eq.\eqref{eq:pert_Einstein_eqn_1} can be approximated as

\begin{equation}
\frac{c^2 \vec{\nabla}^2 \Phi}{a^2} \simeq 4\pi G (\bar{\rho}_m \delta_m + \bar{\rho}_Q \delta_Q) .
\label{eq:pert_Einstein_eqn_1_sub_Hubble}
\end{equation}

\noindent
Now, by substituting Eq.\eqref{eq:pert_Einstein_eqn_1_sub_Hubble} into Eq.\eqref{eq:dm_double_dot_sub_Hubble_1}, we obtain the evolution equation for the matter density contrast as

\begin{equation}
\ddot{\delta}_m + 2 H \dot{\delta}_m - 4\pi G \bar{\rho}_m \delta_m \simeq 4\pi G \bar{\rho}_Q \delta_Q .
\label{eq:dm_double_dot_sub_Hubble_with_de_pert}
\end{equation}

\subsection{Homogeneous dark energy}

In this study, we focus on the homogeneous dark energy models, where dark energy has no fluctuations. Later, in the next study, we shall consider models in which the dark energy is not homogeneous. In the homogeneous dark energy models, we have $\delta_Q=0$ and the evolution equation for the matter density contrast becomes

\begin{equation}
\ddot{\delta}_m + 2 H \dot{\delta}_m - 4\pi G \bar{\rho}_m \delta_m \simeq 0 .
\label{eq:dm_double_dot_sub_Hubble_no_de_pert}
\end{equation}

\begin{figure*}
\centering
\includegraphics[width=0.49\textwidth]{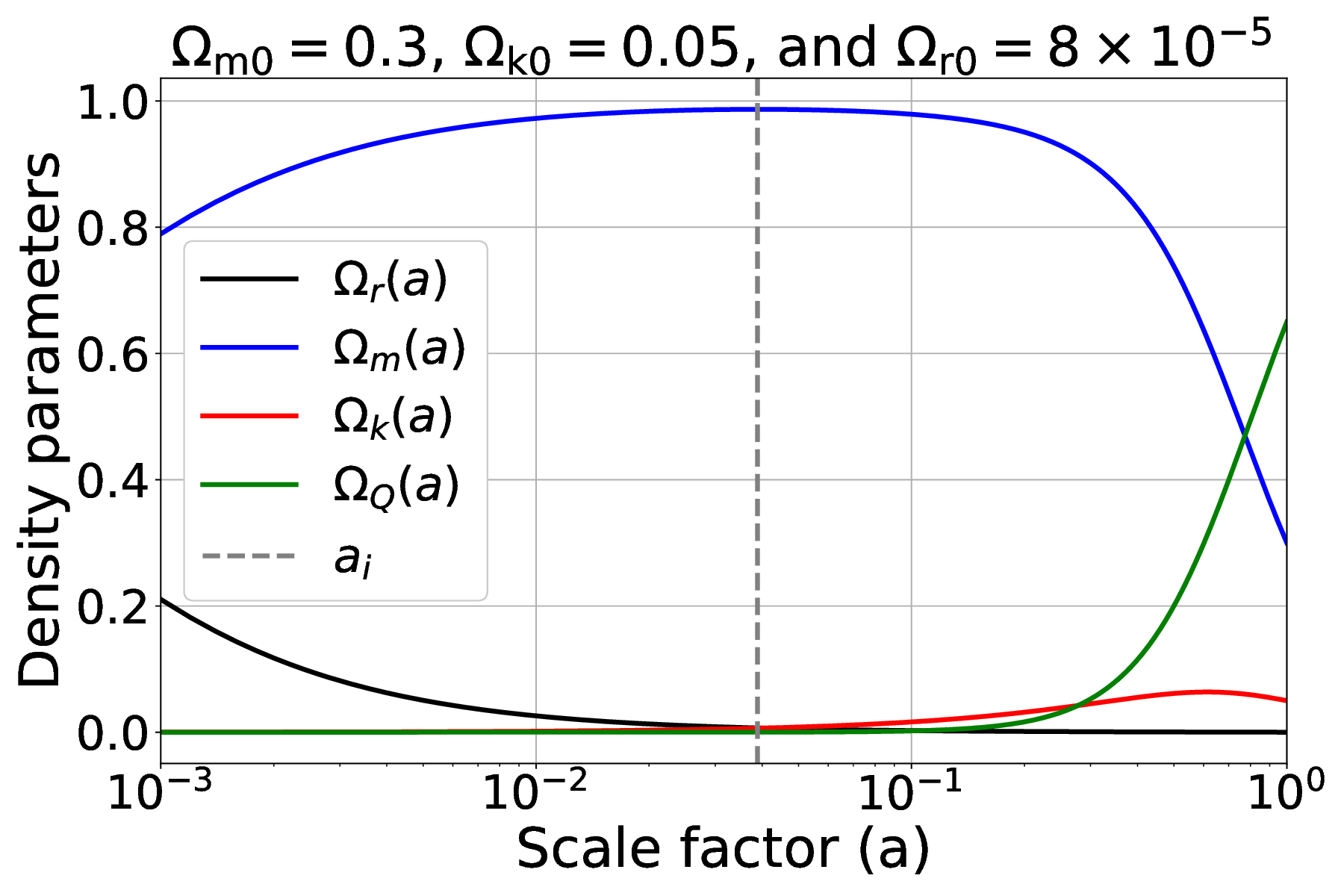}
\includegraphics[width=0.49\textwidth]{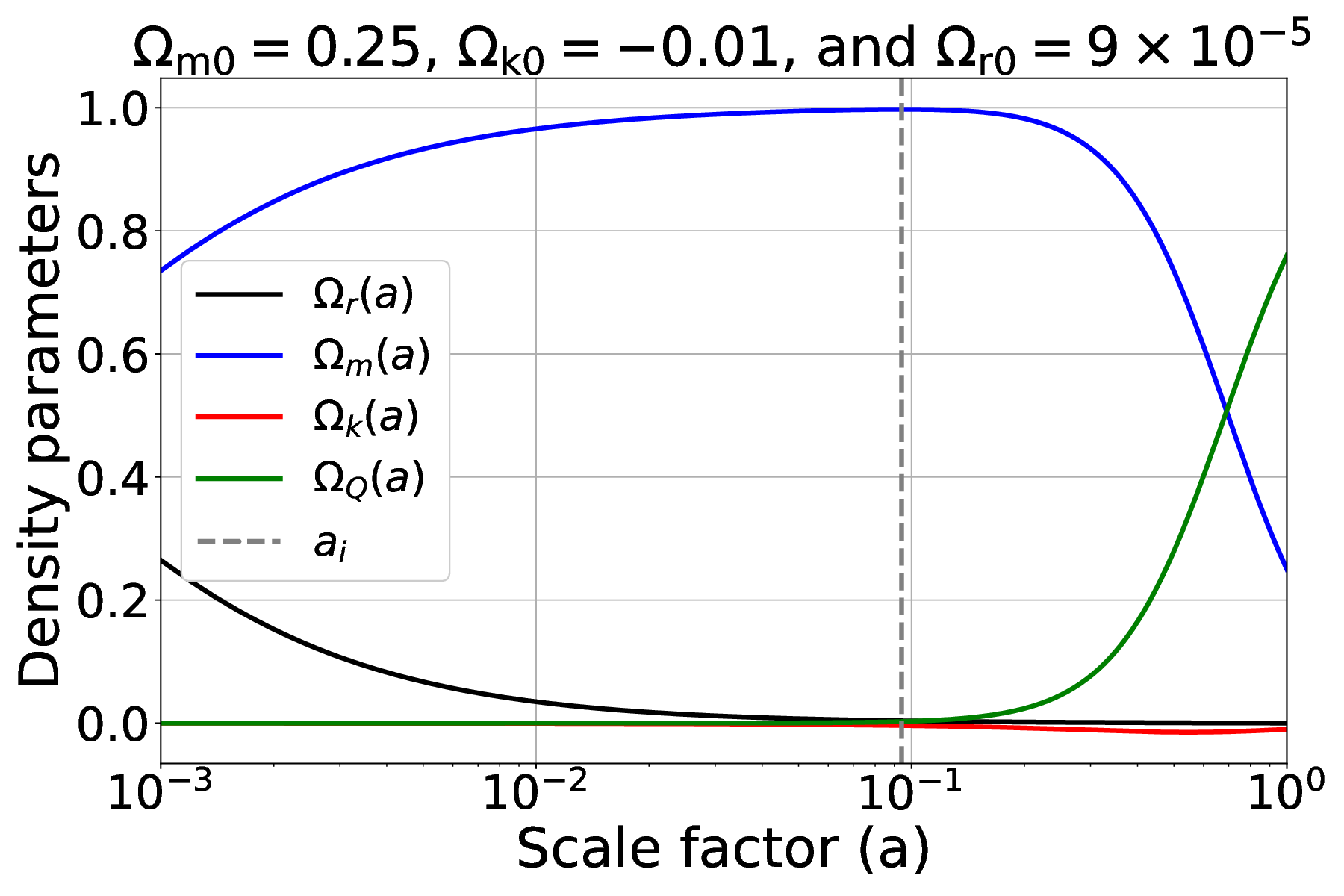}
\caption{
\label{fig:aini}
Energy budget contributions from different constituents. Vertical dashed lines correspond to $a_i$ values.
}
\end{figure*}

\section{Computation of $a_i$}
\label{sec-ai}

We do not choose $a_i$ by hand but compute it using the condition given as

\begin{eqnarray}
\dfrac{d(1-\Omega_m)}{da} \Bigg{|}_{\rm a_i} &=& 0,
\label{eq:compute_ai} \\
\dfrac{d^2(1-\Omega_m)}{da^2} \Bigg{|}_{\rm a_i} &>& 0 . \nonumber
\label{eq:check_minima}
\end{eqnarray}

It is important to note here that when we compute $a_i$ from Eq.\eqref{eq:compute_ai}, we include the contribution of radiation when we define $\Omega_m$ to get the actual peak value of $\Omega_m$ which is closest to the value $\Omega_m \sim 1$. Otherwise, throughout the paper, we do not include radiation because our study is focused only on lower redshifts. In \autoref{fig:aini}, we can see that $\Omega_m \sim 1$ in the scale factor range $10^{-2}$ to $10^{-1}$ (redshift range 10 to 100) and the typical value of $a_i$ we get is of the order of $0.04$ to $0.1$ according to a given model of dark energy and presence or absence of the cosmic curvature. For example, in left panel of \autoref{fig:aini}, we can see $a_i \approx 0.04$ (see vertical dashed line) for the parameter combination $\Omega_{\rm m0}=0.3$, $\Omega_{\rm k0}=0.05$, and $\Omega_{\rm r0}=8 \times 10^{-5}$ in the $\Lambda$CDM model\footnote{Here, $\Omega_{\rm r0}$ is the present value of the radiation energy density parameter.}. In right panel of \autoref{fig:aini}, we consider $\Omega_{\rm m0}=0.25$, $\Omega_{\rm k0}=-0.01$, and $\Omega_{\rm r0}=9 \times 10^{-5}$.

\begin{figure}
\centering
\includegraphics[width=0.49\textwidth]{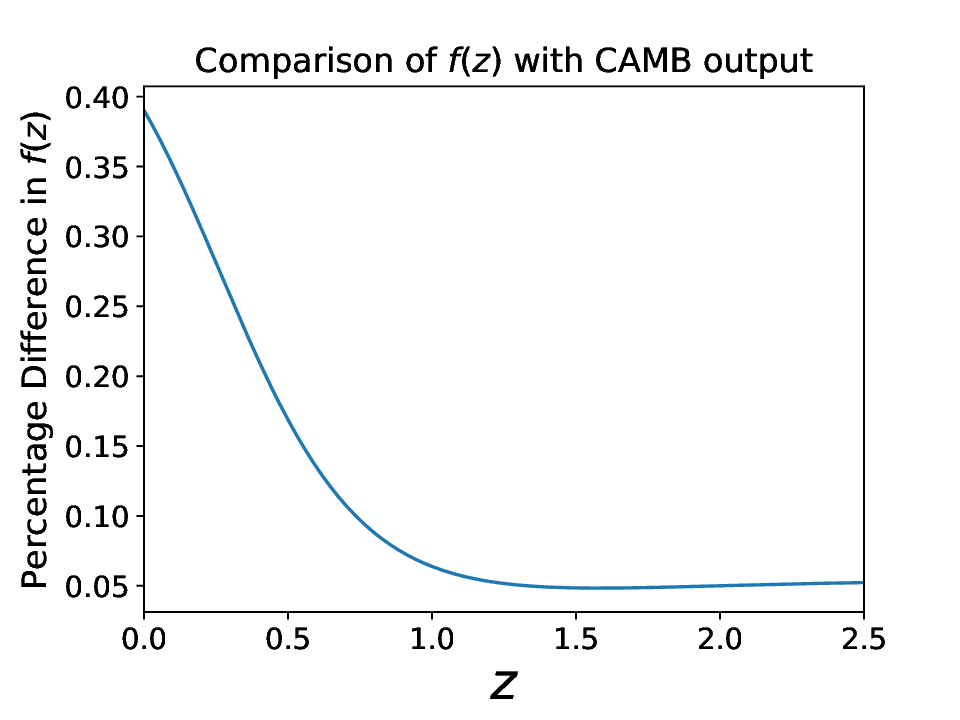}
\caption{
\label{fig:f_cmp}
Percentage comparison in $f$ with CAMB output.
}
\end{figure}

\begin{figure}
\centering
\includegraphics[width=0.49\textwidth]{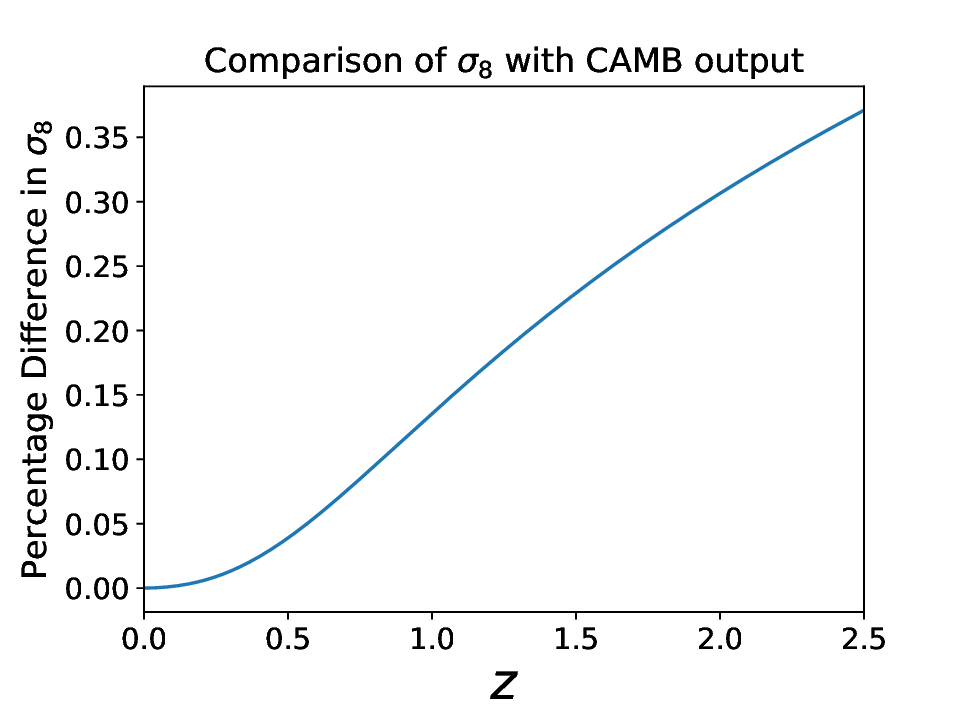}
\caption{
\label{fig:sigma_8_cmp}
Percentage comparison in $\sigma_8$ with CAMB output.
}
\end{figure}

\section{Consistency check with the CAMB}
\label{sec-consistency_camb}

We check the consistency of all the main equations in the main text (both for background expansion and evolution of perturbations) by comparing our results with the CAMB results, where in the CAMB, all the equations are default in general. To compare, we consider a particular combination of parameters as $H_0=70$ km$~$s$^{-1}~$Mpc$^{-1}$, $\Omega_{\rm m0}=0.3$, $\Omega_{\rm k0}=-0.2$, $w_0=-1.5$, $w_a=-0.5$, and $\sigma_{\rm 8,0}=0.8$. In CAMB there are other parameters which we keep to their default values while computing CAMB output. By percentage deviations we mean $\left[\frac{f}{f({\rm CAMB})}-1\right]\times100$ and $\left[\frac{\sigma_8}{\sigma_8({\rm CAMB})}-1\right]\times100$ for $f$ and $\sigma_8$ respectively. The Observables $f$ and $\sigma_8$ are compared in Figs.~\ref{fig:f_cmp} and~\ref{fig:sigma_8_cmp} respectively. Our results are in good agreement with the CAMB results. The differences are well below $1\%$. One can check that the results would be similar for other combinations of parameter values.

\bibliographystyle{apsrev4-1}
\bibliography{references}

\end{document}